\documentclass[useAMS,usenatbib]{mnras}
\usepackage[utf8]{inputenc}

\usepackage{xargs}
\usepackage[pdftex,dvipsnames]{xcolor}
\usepackage{subfigure,wrapfig}
\usepackage[section]{placeins}
\usepackage{graphicx,amsmath,amssymb,multirow,units}
\usepackage{euscript}
\usepackage{zref-savepos}
\usepackage{xspace}
\usepackage{mathptmx}
\usepackage{blindtext}
\usepackage[colorinlistoftodos,prependcaption,textsize=small]{todonotes}
\presetkeys{todonotes}{noline}{}  

\usepackage{breakcites}
\usepackage{microtype}
\usepackage{ctable}

\usepackage{geometry}

\newcommand{\ie}{{i.e.\/}\xspace}
\newcommand{\eg}{{e.g.\/}\xspace}
 
\newcommand{\asec}{\ensuremath{^{\prime\prime}}}
\newcommand{\amin}{\ensuremath{^{\prime}}}
\newcommand{\mjy}{mJy\xspace}

\newcommand{\hatlas}{\textit{H}-ATLAS\xspace}
\newcommand{\lofarhatlas}{LOFAR/\textit{H}-ATLAS\xspace}

\newcommand{\lofar}{\ensuremath{\textrm{LOFAR}}\xspace}
\newcommand{\first}{\ensuremath{\textrm{FIRST}}\xspace}
\newcommand{\spire}{SPIRE\xspace}
\newcommand{\pacs}{PACS\xspace}
\newcommand{\herschel}{\textit{Herschel}\xspace}
\newcommand{\wise}{WISE\xspace}
\newcommand{\sdss}{SDSS\xspace}
\newcommand{\mpajhu}{MPA-JHU\xspace}

\newcommand{\magphys}{{\sc MagPhys}\xspace}

\newcommand{\lofarfreq}{\ensuremath{\mathrm{150\,MHz}}\xspace}
\newcommand{\firstfreq}{\ensuremath{\mathrm{1.4\,GHz}}\xspace}

\newcommand{\kcorrection}{\ensuremath{K}-correction}
\newcommand{\kcorrect}{\ensuremath{K}-correct}
\newcommand{\q}{\ensuremath{q_{250}}\xspace}
\newcommand{\jybeam}{Jy beam$^{-1}$\xspace}
\newcommand{\wisex}{\ensuremath{[4.6] - [12]}\xspace}
\newcommand{\wisey}{\ensuremath{[3.4] - [4.6]}\xspace}

\newcommand\responseauthor[1]{Response:~#1}
\newcommandx{\unsure}[5][1=]{\todo[noprepend,linecolor=red,backgroundcolor=red!25,bordercolor=red,author=#2,caption={#2: #3\newline #4\newline\responseauthor{#5}},#1]{#3}{}}
\newcommandx{\changed}[5][1=]{\todo[noprepend,linecolor=blue,backgroundcolor=blue!25,bordercolor=blue,author=#2,caption={#2: #3\newline #4\newline\responseauthor{#5}},#1]{#3}{}}
\newcommandx{\implemented}[5][1=]{\todo[noprepend,linecolor=OliveGreen,backgroundcolor=OliveGreen!25,bordercolor=OliveGreen,author=#2,caption={#2: #3\newline #4\newline\responseauthor{#5}},#1]{#3}{}}
\newcommandx{\comments}[5][1=]{\todo[noprepend,linecolor=Plum,backgroundcolor=Plum!25,bordercolor=Plum,author=#2,caption={#2: #3\newline #4\newline\responseauthor{#5}},#1]{#3}{}}

\newcounter{whereq}
\newcommand{\location}[1]{\zsaveposy{#1}}
\newcommand{\where}[1]{%
  \stepcounter{whereq}
  \location{whereq-\thewhereq}
  \ifdim\zposy{whereq-\thewhereq}sp<\zposy{#1}sp
    above%
  \else
    below%
  \fi}

\graphicspath{{figures/}}
\DeclareGraphicsExtensions{.pdf}
 
\title[The FIRC at low frequency]{The Far-Infrared Radio Correlation at low radio frequency with \lofarhatlas\thanks{\herschel is an ESA space observatory with science instruments provided by European-led Principal Investigator consortia and with important participation from NASA}}
\author[S.\,C.\,Read et al.]
{S.\,C.\,Read$^1$\,\thanks{E-mail: shaun.c.read@gmail.com}, D.\,J.\,B. Smith$^1$, G.\,Gurkan$^{1,2}$, M.\,J. Hardcastle$^1$, W.\,L. Williams$^1$, P.N.\,Best$^3$, 
\newauthor E. Brinks$^1$, G. Calistro-Rivera$^4$, K.\,T. Chyzy$^5$, K. Duncan$^4$, L. Dunne$^{3,6}$, M.\,J. Jarvis$^{7,8}$, 
\newauthor L.\,K. Morabito$^7$, I. Prandoni$^9$, H.\,J.\,A. R\"ottgering$^4$, J. Sabater$^{3}$, and S. Viaene$^{1, 10}$\\
$^{1}$Centre for Astrophysics Research, School of Physics, Astronomy and Mathematics, University of Hertfordshire, Hatfield, Herts, AL10 9AB, UK\\
$^{2}$CSIRO Astronomy and Space Science, 26 Dick Perry Avenue, Kensington, Perth, 6151, WA, Australia\\
$^{3}$Institute for Astronomy, University of Edinburgh, Royal Observatory, Blackford Hill, Edinburgh, EH9 3HJ, UK\\
$^{4}$Leiden Observatory, Leiden University, P.O. Box 9513, 2300 RA Leiden, The Netherlands\\
$^{5}$Astronomical Observatory, Jagiellonian University, ul. Orla 171, 30-244 Krak\'ow, Poland\\
$^{6}$School of Physics and Astronomy, Cardiff University, The Parade, Cardiff CF24 3AA, UK\\
$^{7}$Astrophysics, University of Oxford, Denys Wilkinson Building, Keble Road, Oxford, OX1 3RH, England\\
$^{8}$Physics and Astronomy Department, University of the Western Cape, Bellville 7535, South Africa\\
$^{9}$INAF-IRA, Via P. Gobetti 101, 40129 Bologna, Italy\\
$^{10}$Sterrenkundig Observatorium, Universiteit Gent, Krijgslaan 281, B-9000 Gent, Belgium}

\begin{document}
\date{\today}

\pagerange{\pageref{firstpage}--\pageref{lastpage}} \pubyear{2018}

\maketitle

\label{firstpage}

\begin{abstract}
  The radio and far-infrared luminosities of star-forming galaxies are tightly correlated over several orders of magnitude; this is known as the far-infrared radio correlation (FIRC).
  Previous studies have shown that a host of factors conspire to maintain a tight and linear FIRC, despite many models predicting deviation. This discrepancy between expectations and observations is concerning since a linear FIRC underpins the use of radio luminosity as a star-formation rate indicator.
  Using \lofar \lofarfreq, \first \firstfreq, and \herschel infrared luminosities derived from the new \lofarhatlas catalogue, we investigate possible variation in the monochromatic (250\,\micron) FIRC at low and high radio frequencies.
  We use statistical techniques to probe the FIRC for an optically-selected sample of 4,082 emission-line classified star-forming galaxies as a function of redshift, effective dust temperature, stellar mass, specific star formation rate, and mid-infrared colour (an empirical proxy for specific star formation rate).
  Although the average FIRC at high radio frequency is consistent with expectations based on a standard power-law radio spectrum, the average correlation at \lofarfreq is not. 
  We see evidence for redshift evolution of the FIRC at \lofarfreq, and find that the FIRC varies with stellar mass, dust temperature and specific star formation rate, whether the latter is probed using \magphys fitting, or using mid-infrared colour as a proxy. 
  We can explain the variation, to within $1\sigma$, seen in the FIRC over mid-infrared colour by a combination of dust temperature, redshift, and stellar mass using a Bayesian partial correlation technique.

\end{abstract}

\begin{keywords}
infrared: galaxies, radio continuum: galaxies, galaxies: star formation
\end{keywords}

\section{Introduction}\label{sec:intro}

The far-infrared luminosities of star-forming galaxies have long been known to correlate tightly and consistently with synchrotron radio luminosity across many orders of magnitude in infrared and radio luminosities, independent of galaxy type and redshift \citep{Kruit1971Observations,Jong1985Radio,Condon1991Correlations,Yun2001Radio,Bell2003Estimating,Bourne2011Evolution}.

The existence of some relation should not be surprising since the basic physics relating emission in each waveband to the presence of young stars is well understood.
Young stars heat the dust within their surrounding birth clouds, which radiate in the infrared \citep{Kennicutt1998Star,Charlot2000Simple}.
The supernovae resulting from the same short-lived massive stars accelerate cosmic rays into the galaxy's magnetic field thereby contributing non-thermal radio continuum emission over $\approx 10^8$ years \citep{Blumenthal1970Bremsstrahlung,Condon1992Radio,Longair2011High}.
However, the fact that the  Far-Infrared Radio Correlation (FIRC) has consistently been found to have low scatter \citep{Helou1985Thermal,Jong1985Radio,Condon1992Radio,Lisenfeld1996FirRadio,Wong2016Determining} is surprising. 
Such tight linearity is consistent with a simple calorimetry model \citep{Voelk1989Correlation}, whereby cosmic ray electrons lose all of their energy before escaping the host galaxy and where all UV photons are absorbed by dust and re-radiated in the infrared. 
This results in synchrotron radiation being an indirect measure of the energy of the electron population and infrared luminosity being proportional to young stellar luminosity. 
Therefore, assuming calorimetry, the ratio of these two measures will remain constant as they are both dependent on the same star formation rate.
The FIRC can therefore be used to bootstrap  a calibration between a galaxy's star formation rate and its radio luminosity \citep[\eg][]{Condon1992Radio,Murphy2011Calibrating} -- but only if there is no additional contribution from AGN.

The physics required to model the FIRC is  complex. For example, the timescale of the electron synchrotron cooling that produces the radio emission is thought to be longer than the timescale for the escape of those electrons \citep{Lisenfeld1996Quantitative,Lacki2010Physics}  for normal spirals, and starlight is only partially attenuated in the UV \citep{Bell2003Estimating}. 
Therefore, it is reasonable to suppose that the calorimetry interpretation must be at least partially inaccurate and that there should be some observable variation in the FIRC over the diverse population of star-forming galaxies. 
In particular, due to their strong magnetic fields, we expected starburst galaxies to be good calorimeters and therefore have a correlation with a slope that is much closer to one than other star-forming galaxies \citep{Lacki2010Physics}.

However, since synchrotron emission depends strongly on magnetic field strength, the assumption about how this changes with galaxy luminosity is crucial to explain the correlation.
Alternatives to the calorimetry model have also been proposed, \eg (i) the model of \citet{Niklas1997New}, where the FIRC arises as the by-product of the mutual dependence of magnetic field strength and star-formation rate upon the volume density of cool gas, and (ii) \citet{Schleicher2016StarForming}, where the FIRC is based on a small-scale dynamo effect that amplifies turbulent fields from the kinetic turbulence related to star formation.
There are a number of reasons to expect the FIRC to vary with the parameters that control synchrotron and dust emission, but it seems that infrared and radio synchrotron must both fail as star formation rate indicators in such a way as to maintain a tight and linear relationship over changing gas density. The model detailed by \citet{Lacki2010Physicsa} and \citet{Lacki2010Physics} suggests that although normal galaxies are indeed electron and UV calorimeters, conspiracies at high and low surface density, $\Sigma_g$, contrive to maintain a linear FIRC.
At low surface density, many more UV photons escape (and therefore lower observed infrared emission) due to decreased dust mass but at the same time, because of the lower gravitational potential, more electrons escape without radiating all their energy, decreasing the radio emission.
Meanwhile, at high surface densities, secondary charges resulting from cosmic ray proton collisions with ISM protons become important \citep{Torres2004Theoretical,Santamaria2005High}. 
Synchrotron emission from those electrons and positrons may dominate the emission from primary cosmic ray electrons.
However, the FIRC is maintained due to the increased non-synchrotron losses from bremsstrahlung and inverse Compton scattering at higher densities. 

These conspiracies rely on fine tuning of many, sometimes poorly known,  parameters in order to balance the mechanisms that control the linearity of the FIRC. If we expect variation over star-forming galaxies due to differences in gas density, stellar mass, and redshift (to name a few), then we should probe the FIRC over known star-forming sequences such as those found in colour-magnitude \citep{Bell2004Nearly} and mid-infrared colour-colour diagrams \citep[\eg][]{Jarrett2011SpitzerWise,Coziol2015Comparing}, and the star formation rate -- stellar mass relation \citep{Brinchmann2004Physical,Noeske2007Star,Peng2010Mass,Rodighiero2011Lesser}.

Naively, we might also expect some variation of the FIRC with redshift. 
At the very least, radio luminosity should decrease with respect to infrared luminosity due to inverse Compton losses from cosmic microwave background (CMB) photons \citep{Murphy2009FarInfraredRadio}.
The CMB energy density increases proportional to $(1+z)^4$ \citep{Longair1994High}, so the ratio of infrared to radio luminosity should noticeably increase with redshift even at relatively local distances, assuming a calorimetry model and that CMB losses are significant.

However, this is one of the key areas of dispute between different observational studies. While the many works find no evidence for evolution \citep[\eg][]{Garrett2002FirRadio,Appleton2004Far,Seymour2009Investigating,Sargent2010VlaCosmos}, there are exceptions \citep[\eg][]{Seymour2009Investigating,Ivison2010FarInfraredRadio,Michaowski2010Rapid,Michaowski2010Cosmic,Basu2015RadioFarInfrared,Rivera2017Lofar,Delhaize2017VlaCosmos}. 
Particular among those studies, \citet{Rivera2017Lofar} find a significant redshift trend at both \lofarfreq and \firstfreq when using the \textit{Low Frequency Array} \citep[\lofar,][]{Haarlem2013Lofar} data taken over the Bo{\"o}tes field.
The FIRC has been studied extensively at \firstfreq \citep{Jong1985Radio,Condon1991Correlations,Bell2003Estimating,Jarvis2010HerschelAtlas,Bourne2011Evolution,Smith2014Temperature} but rarely at lower frequencies. 
These low frequencies are particularly important, since new radio observatories such as \lofar are  sensitive in the $15-200\,\mathrm{MHz}$ domain, where at some point the frequency dependence of optical depth results in the suppression of synchrotron radiation by free-free absorption \citep{Schober2017Tracing}, causing the radio SED to turn over. 
As a result, there will be some critical rest-frame frequency below which we can expect a substantially weaker correlation between a galaxy's radio luminosity and its star formation rate.\footnote{This frequency at which a galaxy's radio SED turns over will depend heavily upon gas density and ionisation, and so we expect it to vary from galaxy to galaxy.} 
Moreover, at the higher frequencies probed by \textit{Faint Images of the Radio Sky at Twenty centimetres} \citep[\first,][]{Becker1995First} (\firstfreq), there may be a thermal component present in the radio emission \citep{Condon1992Radio}, which tends to make the correlation between infrared and higher radio frequencies more linear. 
However, due to the poor sensitivity of \first to star-forming galaxies with low brightness temperatures (galaxies with $T_{\textrm{bright}}<10\mathrm{K}$ will not be detected by \first), we cannot expect the  thermal components of detected sources to help linearise the FIRC at \firstfreq.
At low frequencies, these effects become less important and so the perspective they provide is useful in disentangling the effect of thermal contributions and lower luminosity galaxies on the FIRC.
Given the potential ramifications for using low-frequency radio observations as a star formation indicator, this possibility must be investigated.  
Indeed, \citet{Gurkan2018LofarHAtlas} have found that a broken power-law is a better calibrator for radio continuum luminosity to star-formation rate, implying the existence of some other additional mechanism for the generation of radio-emitting cosmic rays.

Furthermore, lower radio frequencies probe lower-energy electrons, which take longer to radiate away their energy than the more energetic electrons observed at 1.4\,GHz, and this results in a relationship between the age of a galaxy's electron population and the radio spectral index  \citep{Scheuer1968Radio,Blundell2001Spectral,Schober2017Tracing}.
Therefore, even if the FIRC is linear at high frequencies due to some conspiracy, this will not  necessarily be the case at low frequencies.
An investigation of the FIRC at low frequency will test models of the FIRC which rely on spectral ageing to maintain linearity \citep[\eg][]{Lacki2010Physics}. 

Combined with the fact that radio observations are impervious to the effects of dust obscuration, this makes low-frequency radio observations a very appealing means of studying star formation in distant galaxies, providing that the uneasy reliance of SFR estimates on the FIRC can be put on a more solid footing. 
The nature of the FIRC conspiracies varies over the type of galaxy and its star formation rate \citep{Lacki2010Physicsa}.
The detection of variation in the FIRC over those galaxy types, or lack thereof, will provide important information about the models that have been constructed \citep[\eg][]{Lacki2010Physicsa,Schober2017Tracing}. 
Several methods are used to distinguish galaxy types for the purposes of studying the FIRC, particularly to classify these into star-forming galaxies and AGN such as BPT diagrams \citep{Baldwin1981Classification}, panchromatic SED-fitting with AGN components \citep{Berta2013Panchromatic,Ciesla2016Imprint,Rivera2016Agnfitter}, and classification based on galaxy colours.
Among these, galaxy colours provide a readily accessible method to distinguishing galaxy types or act  as proxies for properties such as star formation rate.
Diagnostic colour-colour diagrams are commonplace in galaxy classification; infrared colours in particular have been widely used to distinguish between star-forming galaxies and AGN \citep{Lacy2004Obscured,Stern2005MidInfrared,Jarrett2011SpitzerWise,Mateos2012Using,Coziol2015Comparing}. 
In order to investigate the potential difference in the FIRC over normal galaxies as well as in starbursts we  use the mid-infrared diagnostic diagram, (MIRDD, \citealt{Jarrett2011SpitzerWise}) .
Constructed from the \textit{Wide-field Infrared Survey Explorer} \citep[\wise,][]{Wright2010WideField} \wisex and \wisey colours, SWIRE templates \citep{Polletta2006Chandra,Polletta2007Spectral} and GRASIL models \citep{Silva1998Modeling} can be used to  populate the MIRDD with a range of galaxy types spanning a redshift range of $0 < z < 2$. This MIRDD not only distinguishes AGN and SFGs but also describes a sequence of normal star-forming galaxies whose star formation rate increases to redder colours. 

Past \firstfreq surveys such as \first and the \textit{NRAO VLA Sky Survey} \citep[NVSS, ][]{Condon1998Nrao} have been extremely useful in studying star formation, though there are inherent problems in using them to do this. 
NVSS is sensitive to extended radio emission on the scale of arcminutes. 
However, its sensitivity of $\sim 0.5$ m\jybeam and resolution of 45\,\asec  means that it has trouble identifying radio counterparts to optical sources  and its flux limit means that it will peferentially detect bright or nearby sources. 
\first has both a higher resolution and a higher sensitivity than NVSS (5\,\asec with $\sim 0.15$ m\jybeam). 
However, due to a lack of short baselines, \first resolves out the extended emission frequently present in radio-loud AGN and in local star-forming galaxies \citep{Jarvis2010HerschelAtlas}. 
This makes it difficult to remove galaxies dominated by AGN and to directly compare star-forming galaxies over different wavelengths. 
Meanwhile, \lofar offers the best of both worlds: a large field of view coupled with high sensitivity on both small and large scales  and high resolution \citep{Haarlem2013Lofar} at frequencies between 30 and 230\,MHz. 
Operating at \lofarfreq,  \lofar contributes a complementary view to the wealth of data gathered at higher frequencies. 
The sparsely examined low-frequency regime offered by \lofar combined with its increased sensitivity and depth relative to other low-frequency instruments allows us to probe the FIRC in detail, and to test predictions of its behaviour relative to relations at higher frequencies that we measure with \first.

This study will analyse the nature of the FIRC at low and high frequencies and over varying galaxy properties. 
How does the FIRC evolve with redshift? Does it vary as a function of \wise mid-infrared colour? 
Do the specific star-formation rate (as fit by \magphys) and stellar mass impact these questions?
We answer these questions for our data set and compare these metrics with those found at higher frequencies and with literature results using different selection criteria.

This work uses the same base dataset as \citet{Gurkan2018LofarHAtlas}.
The same aperture-corrected fluxes extracted from \herschel , \lofar, and \first images are used here.
Our investigation differs from theirs in that we concentrate on the observed variation of the FIRC over dust properties whereas \citet{Gurkan2018LofarHAtlas} focus on the direct characterisation of radio star-formation rates.
In Section~\ref{sec:data}, we describe our data sources and the method of sample selection. 
In Section~\ref{sec:methods} we outline our methods for calculating \kcorrection s, luminosities, and the methods used to characterise the variation of the FIRC. 
We present and discuss the results of these procedures in Section~\ref{sec:results}, and summarise our conclusions in Section~\ref{sec:conclusions}. 

We assume a standard $\Lambda$CDM cosmology with $H_0 = 71$\,km\,s$^{-1}$\,Mpc$^{-1}$, $\Omega_M = 0.27$ and $\Omega_\Lambda = 0.73$ throughout, and for consistency with \citet{Jarrett2011SpitzerWise}, all magnitudes are in the Vega system.

\section{Data Sources}\label{sec:data}The dataset we use here is the same as \citet{Gurkan2018LofarHAtlas}  in that the infrared and radio aperture-corrected fluxes are drawn from the same catalogue.
However, due to two effects listed below, our star-forming sample is selected using a different method.
Firstly, a potential contamination of AGN will have a large effect on the detected variation of infrared-to-radio luminosity ratio  over mid-infrared colours. 
We therefore require stronger signal-to-noise criteria ($5\sigma$ detections in the BPT optical emission lines)  than the one in use in \citet{Gurkan2018LofarHAtlas} ($3\sigma$). 
Secondly, using the \citet{Gurkan2018LofarHAtlas} star-forming selection criterion but with a $5\sigma$ requirement results in too few star-forming galaxies with reliable $5\sigma$ detections in the first three \wise bands. 
In order to increase our sample size but maintain robust classification we employ methods detailed below. 

\subsection{Sample selection}
\label{sec:data:sample_selection}

To avoid introducing a possible bias by selecting our sample from far-infrared and\slash or radio catalogues, our sample is drawn from the \mpajhu catalogue \citep{Brinchmann2004Physical} over the region of the \textit{North Galactic Pole} (NGP) field covered by the \lofarhatlas\ survey, which is described in sections \ref{sec:data:hatlas} and \ref{sec:data:radio}.
The \mpajhu catalogue uses an optimised pipeline to re-analyse all \sdss \citep{York2000Sloan} spectra, resulting in a sample with reliable spectroscopic redshifts, improved estimates of stellar mass, and star formation rate, as well as emission line flux measurements for each galaxy.
We use their latest analysis performed on the \sdss DR7 release \citep{Abazajian2009Seventh} to obtain optical emission line fluxes and spectroscopic redshifts for \kcorrection s. 

To select our star-forming sample, we first obtain all optically selected 15,003 sources in the MPA-JHU catalogue with reliable (\texttt{ZWARNING = 0}) spectroscopic redshifts $z < 0.7$ in the region covered by our \lofarhatlas\ data. 
Since we are interested in studying the FIRC, we wish to focus only on star-forming galaxies, and remove those sources with evidence for contamination by emission from an active galactic nucleus (AGN). 
Our priority is to seek an unbiased sample at the cost of such a sample not necessarily being complete.
We do this using the BPT \citep{Baldwin1981Classification} emission line classification method, requiring fluxes detected at $\ge 5\sigma$ in $H\alpha$, $H\beta$, $[OIII]\lambda5007$, and $[NII]\lambda6584$, together with the star-forming/composite line defined by \citet{Kewley2001Optical}. 
3,082 galaxies, with redshifts $z < 0.4$, are identified as star-forming in this manner.  

\begin{figure}
  \centering
  \includegraphics[width=\linewidth]{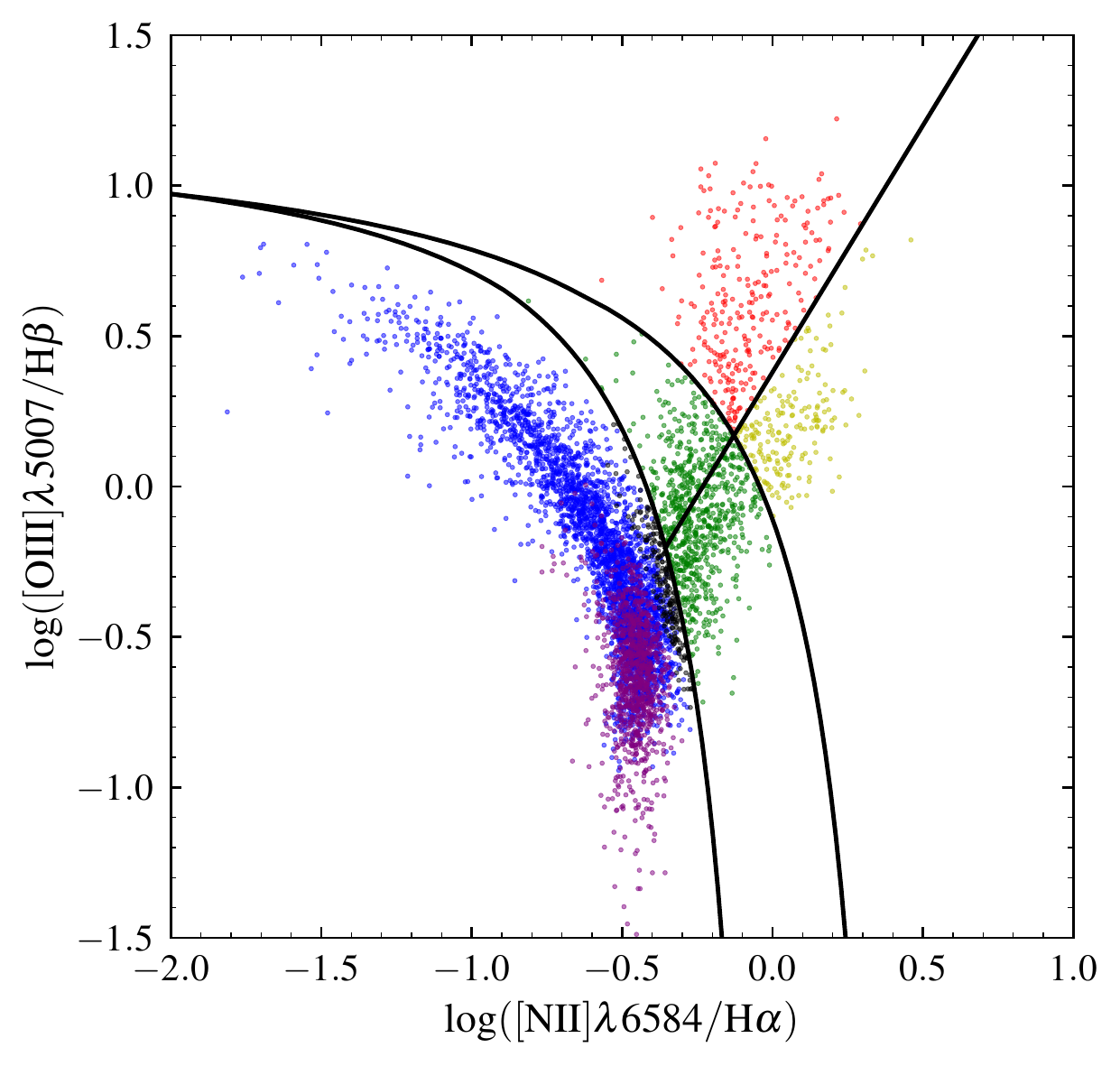}
  \caption{Emission line ratio diagnostic diagram. The coloured points represent Seyfert 2s (in red), star-forming galaxies (blue), transition objects (green), and LINERs (yellow). 
   The black points are those galaxies whose $5\sigma$ upper limit on $[OIII]\lambda5007$ flux would not classify them as purely star-forming (not included in our sample).
   The purple points show those additional galaxies whose upper limits in $[OIII]\lambda5007$ still classifies them as star-forming.
  The upper and lower solid black lines used to distinguish between populations are from \citet{Kewley2001Optical} and \citet{Kauffmann2003Host} respectively.}
  \label{fig:bpt}
\end{figure}

To give us the largest possible sample of star-forming galaxies, we  include those galaxies  with $5\sigma$ detections in $[NII]\lambda6584$, $H\alpha$, and $H\beta$, provided that the upper limit on the $[OIII]\lambda5007$ flux  in the MPA-JHU catalogue enables us to unambiguously classify them as star-forming.
By using this method, we can be sure that they lie below the star-forming/composite line from \citet{Kewley2001Optical} in Figure~\ref{fig:bpt}.
We identify an additional 1,012 star-forming galaxies using this criterion, and they are shown in purple in Figure~\ref{fig:bpt}. 
In addition, we remove the 12 sources which lie within the QSO box defined in \citet{Jarrett2011SpitzerWise}.
This provides us with our main sample of 4,082 star-forming galaxies with $z < 0.4$  for use in comparing the FIRC at high and low frequencies.
We constructed the MIRDD \citep{Jarrett2011SpitzerWise} based on \textit{WISE All Sky Survey} \citep[\wise,][]{Cutri2012Vizier} fluxes (with no \kcorrection\xspace applied) to identify the location of our galaxies compared to a range of sources of different types. Since we are binning  across \wise colour spaces, we construct a second sample for the mid-infrared  analysis only, requiring 5$\sigma$ detections in the first three \wise bands  (centred on 3.4\,\micron, 4.6\,\micron, and 12\,\micron). 
This results in a sub-sample of 2,901 sources for use in  tracing the FIRC over the mid-infrared colour space depicted in Figure~\ref{fig:wise_sample}.
Our sample sizes are shown in Table~\ref{tab:samples}.

\begin{table}
\caption{Number of star-forming galaxies within each sub-sample detected with \herschel at 250\,\micron~, \lofar at \lofarfreq, and \first at \firstfreq.}
\begin{tabular}{llll}
\toprule
                &          & 1. All SFGs & 2. WISE detected SFGs \\
\midrule
    $>3\sigma$ & \herschel &        3351 &                  2673 \\
                  & \lofar &        2436 &                  2016 \\
                  & \first &        1438 &                  1098 \\
        & both radio bands &        1008 &                   863 \\
\midrule
\midrule
    $>5\sigma$ & \herschel &        2616 &                  2209 \\
                  & \lofar &        1876 &                  1627 \\
                  & \first &         835 &                   640 \\
        & both radio bands &         533 &                   455 \\
\midrule
total & {} &                        4082 &                  2901 \\
\bottomrule
\end{tabular}
\label{tab:samples}
\end{table}

We do not use the catalogue of detected sources summarised by Table~\ref{tab:samples} for our analysis here. 
Such a catalogue will inevitably become contaminated with noise spikes. 
Instead, we employ averaging techniques described below in order to treat non-detections and detections in the same manner.
We don't make any signal-to-noise cuts beyond those imposed on the BPT emission lines used in the star-forming classification.
In addition, our samples are drawn from the \mpajhu catalogue and so this imposes a strong optical prior on the location of a given source.
This allows us to conduct forced aperture photometry, in order to estimate radio fluxes (see Section~\ref{sec:data:photometry}), for our entire sample with a high degree of confidence that the aperture is correctly placed.

\subsection{Infrared data}
\label{sec:data:hatlas}

The far-infrared data used in this study come from the \hatlas\ survey \citep{Eales2010Herschel,Valiante2016HerschelAtlas,Smith2017Herschel,Maddox2018Herschel,Furlanetto2018Second}. 
\hatlas\ is the largest extragalactic {\it Herschel} survey, covering a total of 510 deg$^2$ in five infrared bands with the \textit{Photoconductor Array Camera and Spectrometer} \citep[\pacs,][]{Ibar2010HAtlas,Poglitsch2010Photodetector} and \textit{Spectral and Photometric Imaging Receiver} \citep[\spire,][]{Griffin2010HerschelSpire,Pascale2011First,Valiante2016HerschelAtlas} instruments (sampling wavelengths of 100, 160, 250, 350, and 500\,\micron). 
The \hatlas catalogues have a 5$\sigma$ noise level of 33.5\,\mjy at 250\,\micron, which is the most sensitive band \citep{Ibar2010HAtlas, Rigby2011HerschelAtlas,Smith2011HerschelAtlas,Smith2012Herschel,Smith2017Untitled}. 
In this study, we focus on the \hatlas\ observations covering 142\,deg$^2$ of the NGP field.

\subsection{\lofar data from \lofarhatlas}
\label{sec:data:radio}

\lofar has observed the \hatlas NGP field at the sensitivity and resolution of the \lofar Two-Metre Sky Survey (LoTSS, \citealt{Shimwell2017Lofar}, Williams et al. in prep., Duncan et al. in prep.). 
Whilst the first implementation of the \lofarhatlas survey \citet{Hardcastle2016LofarHAtlas} used a facet-calibration technique, this paper uses data calibrated by a significantly improved method. 
The new direction-dependent calibration technique uses the methods of \citet{Tasse2014Applying,Tasse2014Nonlinear}. 
The calibrations are implemented in the software package \textsc{KILLMS} and imaged with \textsc{DDFACET} \citep{Tasse2018Faceting} which is built to apply these direction-dependent calibrations.
The \lofarhatlas data were processed using the December 2016 version of the pipeline, \textsc{DDF-PIPELINE2}\footnote{See \url{http://github.com/mhardcastle/ddf-pipeline} for the code.} \citep[][and in prep.]{Shimwell2017Lofar}. 
This reprocessing yields a higher image fidelity and a lower noise level than the process detailed by \citet{Hardcastle2016LofarHAtlas}.
It not only increases the point-source sensitivity and removes artefacts from the data, but also allows us to image at (slightly) higher resolution. 
The images used here (as in \citealt{Gurkan2018LofarHAtlas}) have a restoring beam of 6 arcsec FWHM, and 50 per cent of the newly calibrated \lofarhatlas field has an RMS below $\sim0.25$ m\jybeam and 90 per cent is below $\sim0.85$ m\jybeam.

\subsection{Photometry}\label{sec:data:photometry}
Since we used optical data to select our sample, flux limited catalogues from the \lofar, \first, or \hatlas surveys do not contain photometry for every source in our sample, since some of our sources are not formally detected (\eg to $\ge 5\sigma$). 
Moreover, some sources are larger than the \herschel beam and so matched filter images are not preferred.
Instead, the dataset used here (from \citealt{Gurkan2018LofarHAtlas}) follows \citet{Jarvis2010HerschelAtlas}, \citet{Smith2014Temperature}, and \citet{Hardcastle2016LofarHAtlas}, by measuring \lofar, \first, and \herschel flux densities using forced aperture photometry. 

In order to have consistent flux densities across radio and infrared bands, we use 10 arcsec radius circular apertures, centred on each source's optical position, finding that this size of aperture is optimal since it is small enough to limit the influence of confusion noise, and large enough to mean that aperture corrections are small.
The uncertainties on both \lofar and \first flux densities were estimated using their respective r.m.s.\ maps: scaling the noise value in the image at the pixel coordinate of each source by the square root of the number of beams in the aperture.
We do not correct for thermal contributions, whereby the thermal SED also contributes at radio frequencies, in \first or \lofar.
In the \herschel bands, we add the recommended calibration uncertainties in quadrature ($5$ per cent for \pacs and $5.5$ per cent for \spire) \citep{Valiante2016HerschelAtlas,Smith2017Herschel}. 

\begin{figure}
  \centering
  \includegraphics[width=\linewidth]{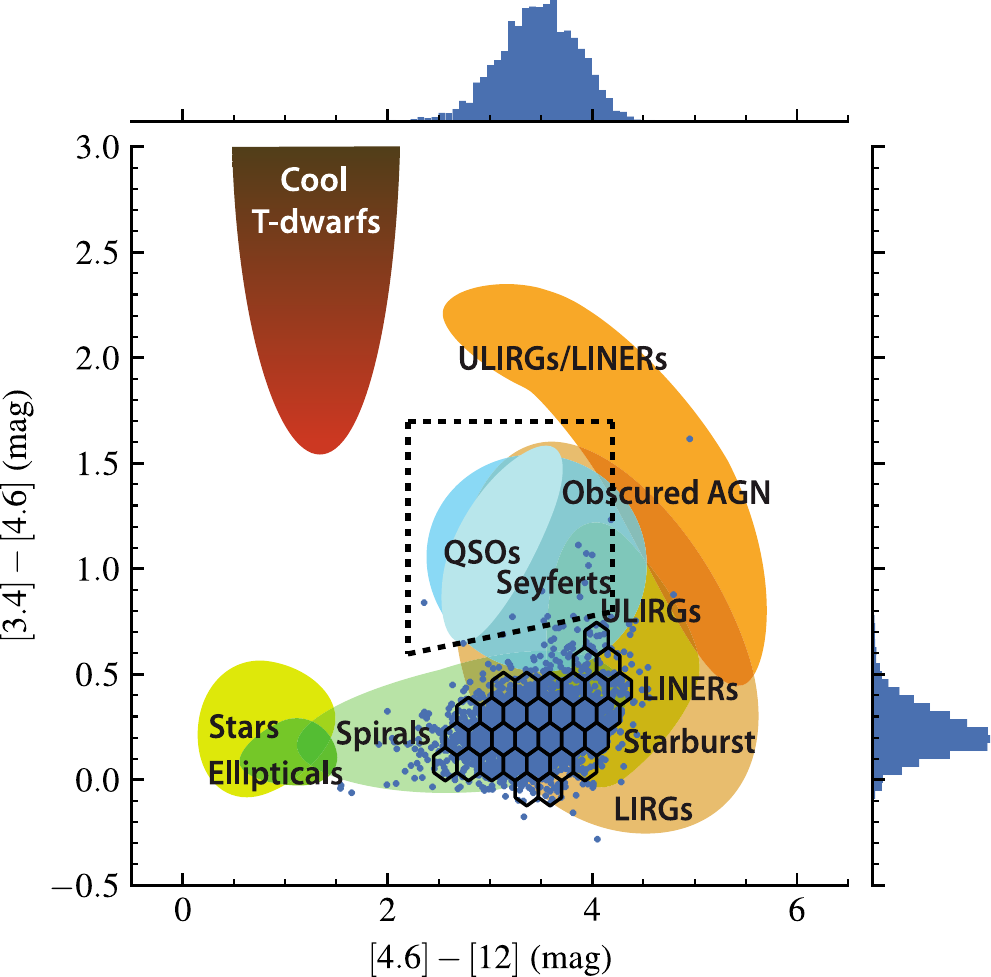}
  \caption{The mid-infrared sub-sample ($5\sigma$ WISE detections, shown as blue points) overlaid on the \citet{Jarrett2011SpitzerWise} MIRDD which uses the magnitudes at three WISE wavelengths $W1$ at 3.4\,\micron, $W2$ at 4.6\,\micron, and $W3$ at 12\,\micron. 
  The coloured regions are as published in \citet{Wright2010WideField}, and intended to show the approximate locations of galaxies of a range of different types. 
  The hexagonal bins over the region centred on $[4.6]-[12] \approx 3.5$, and $[3.4]-[4.6] \approx 0.25$ are used to trace \q in later sections of this paper, and are shown here to provide context.
  The QSO box defined by \citep{Jarrett2011SpitzerWise} is depicted as a dashed box. 
  Number counts over both colours are shown as blue histograms.
  }\label{fig:wise_sample}
\end{figure}

\section{Methods}\label{sec:methods}
\subsection{Low frequency luminosities}
\label{sec:methods:radio}

We calculate $K$-corrected \lofarfreq luminosity densities for every source in our sample assuming that $S_\nu \propto \nu^{\alpha}$, with a spectral index of $-0.71$ \citep{Condon1992Radio,Mauch2013325Mhz}:

\begin{equation}
  L_{\nu} = 4\pi d_L^{2}(z) S_{\nu, obs} (1+z)^{-\alpha - 1},
\label{eq:radio luminosity}
\end{equation}

\noindent where the additional factor of $(1+z)^{-1}$  accounts for the bandwidth correction, and $d_L(z)$ is the luminosity distance in our adopted cosmology.

There is an additional uncertainty on the $K$-corrected luminosity densities due to assuming a constant spectral index; we attempt to account for this by bootstrapping based on the \citet{Mauch2013325Mhz}  distribution of star-forming spectral indices. 
For each galaxy we draw 1000 spectral indices from the prior distribution centred on $-0.71$ with an RMS of 0.38. 
The luminosity densities are calculated using Equation~\ref{eq:radio luminosity} with uncertainties estimated based on the standard deviation of the bootstrapped distribution, however we note that the \kcorrection s and their uncertainties derived for our sample are small since all sources are below $z=0.4$.

\subsection{Far-infrared luminosities}
\label{sec:methods:ir}To estimate the intrinsic far-infrared luminosity densities, we assume an optically thin greybody for the dust emission:

\begin{equation}\label{eq:greybody}
  S_{\nu} \propto \frac{\nu^{3+\beta}}{\exp{(h\nu / kT)} - 1},   
\location{eq:greybody}
\end{equation}

\noindent where $T$ is the dust temperature, $k$ is the Boltzmann constant, $h$ is the Planck constant, and $\beta$ is the emissivity index.  
The dust emissivity varies as a power law over frequency and its inclusion as the constant $\beta$ attempts to summarise the varying dust compositions into a single galaxy-wide isothermal component. 
Taking $\beta = 1.82$ has been found to provide an acceptable fit to the infrared SEDs of galaxies in the \hatlas\ survey \citep{Smith2013Isothermal} and so we assume the same value for $\beta$ here.
We fit Equation~\ref{eq:greybody} to the \herschel \pacs/\spire fluxes at 100, 160, 250, 350, and 500\,\micron. 
We include the \pacs wavelengths despite their reduced sensitivity since they have been found to be important in deriving accurate temperatures \citep{Smith2013Isothermal}. 

We use the Python package \textsc{emcee} \citep{Mackey2013Emcee} which is an implementation of the \citet{Goodman2010Ensemble} Affine Invariant MCMC Ensemble Sampler (AIMCMC). 
AIMCMC is known to sample from degenerate and highly correlated posterior distributions with an efficiency superior to traditional Metropolis techniques \citep{Goodman2010Ensemble}.
For each galaxy, 10 walkers are placed at initial temperatures drawn from a prior normal distribution centred at 30K with a standard deviation of 100K. 
We find that altering the width of the temperature prior does not affect our results.

The walkers sample the probability distribution set by the least squares likelihood function. 
At each temperature that the walkers sample, the resultant grey-body is redshifted to the observed frame and propagated through the \herschel response curves.
We ran the sampler for 500 steps with the 10 walkers and a burn-in phase of 200 steps.
Each galaxy therefore has 3000 informative samples to contribute to the probability distributions.
In addition to the temperature for each MCMC step, we recorded the modelled intrinsic luminosity densities, modelled observed fluxes, and \kcorrection s for each each infrared wavelength. 
This allowed us to find the probability distributions for these parameters and hence their uncertainties in a Bayesian manner.

\subsection{Calculating the FIRC}\label{sec:methods:firc}
The FIRC is traditionally parametrised by the log of the ratio of infrared to radio luminosity, $q$ \citep{Helou1985Thermal,Bell2003Estimating,Ivison2010FarInfraredRadio}. 
However, the lack of \pacs 60\,\micron~coverage and small number of sources ($< 5 $ per cent) with \wise 22\,\micron~fluxes in the \hatlas NGP field prohibits an accurate estimation of $q$ based on total dust luminosity for a statistically significant sample. Therefore, we calculate a \kcorrect ed monochromatic  \q in the \spire 250\,\micron~band following \citet{Jarvis2010HerschelAtlas} and \citet{Smith2014Temperature}.
\begin{equation}
  q_{250} = \log_{10} \left(\frac{L_{250}}{L_{\rm{rad}}}\right)
\end{equation}The uncertainties on our monochromatic \q estimates are found by propagating uncertainties from the $K$-corrected luminosity densities in the radio and 250\,\micron.
We note that in all of the following sections, we calculate \q using $L_{rad}$ calculated at 150\,MHz in the rest-frame.

In addition to the individual \q found for each galaxy we use a stacking method to evaluate trends across colour spaces, redshift, and temperature.
Averaging \q is fraught with problems such as underestimation caused by AGN contamination, undesirable influence by outlier sources, and amplification of those effects by using the average of the ratio of luminosity densities rather than the ratio of the average luminosity densities (luminosity stacking). 
To make matters worse, selection in either the radio or infrared band used to evaluate the FIRC introduces an inherent SED related bias \citep{Sargent2010VlaCosmos}. 
Here we have mitigated the effects of such biases by selecting in an independent optical band.
To mitigate the effect of outliers and AGN, the ratio of the median luminosity densities has previously been used \citep[\eg][]{Bourne2011Evolution,Smith2014Temperature}.
Median averaging is sometimes preferred since it is more resistant to outliers (\eg residual low-luminosity AGN which may not have been identified by the emission line classifications), and since the median often remains well-defined even in the case of few individual detections \citep[\eg][]{Gott2001Median}. 
However, the distributions of luminosity density even in finite-width bins of redshift are skewed.
We find that a median-stacked \q calculated for the whole star-forming sample does not agree with the likewise-stacked \q in bins of redshift (in that the median of the medians is not close the global median -- this is not the case with the mean).
If we use the mean-stacked \q, we arrive at an agreement between the global and binned \q across redshift.
Due to this counter-intuitive disagreement between measures of \q and the importance of being able to quantify a change in the FIRC over redshift, we use the ratio of the mean luminosity densities (mean-stacked) to evaluate \q.
Although we may side-step issues regarding skewed distributions by using the mean, we are now potentially more affected by outliers and AGN.
We will discuss the possible influence of AGN on our results in more detail in the coming sections. 

To calculate our mean \q values, we use a method similar to \citet{Smith2014Temperature} and take the quotient of the mean radio and 250\,\micron~luminosity density for each bin.
Uncertainties are estimated on each stacked \q using the standard deviation of the distribution resulting from re-sampling this mean 10,000 times with replacement (bootstrapping).
This bootstrapped uncertainty of \q is representative of the distribution of the luminosity densities being stacked. 
To complement the parametrisation of the FIRC with \q, we also fit the FIRC as a power-law with finite intrinsic width \footnote{The intrinsic width of the power-law fit, $\log[\sigma]$, is defined as the logged fractional width of the Gaussian over the power-law line, which we define as: $L_{150\,\mathrm{MHz}} \sim (k L_{250\,\micron}^{\gamma}) (1+\epsilon)$, where $\epsilon \sim \mathcal{N}(0, \sigma)$. 
We fit the parameter $\sigma$ along with $\gamma$ and $k$ in our MCMC run.} to the data using Equation~\ref{eq:slope}
\begin{equation}
  L_{\mathrm{radio}} = kL_{250}^{\gamma},
  \label{eq:slope}
\end{equation}
where $k$ is the normalisation and $\gamma$ is the slope of the FIRC. 
We take into account non-detections by re-sampling from each data point's uncertainty and discarding the negative-value realisations.
We use \textsc{emcee} to fit the power-law with 6000 steps and 32 walkers.
Fitting a power-law allows us to probe the physical mechanisms of radio continuum emission generation.
A value of the slope close to one indicates that the conditions required for calorimetry are satisfied and the FIRC is linear. 
A super-linear slope might result from an escape-dominated scenario whereby cosmic rays escape before emitting in the radio.
At sub-linear slopes, losses from cooling processes such as inverse-Compton dominate \citep{Li2016ChangEs}.

We have discussed two methods of quantifying the FIRC (mean-stacked \q and power-law fit). 
In addition, there are three types of uncertainty in the FIRC that we discuss in this analysis: 
\begin{enumerate}
  \item The uncertainty in \q, calculated as the width of the bootstrapped distribution of stacked \q.
  \item The uncertainty in the slope of the FIRC, $\gamma$, quantified by MCMC fit.
  \item The change in stacked \q, $\gamma$, and other statistical results due to the presence of misclassified AGN. 
\end{enumerate}
We estimate the change in our results due to misclassified AGN in Section~\ref{sec:results:agn} where we run our analysis again, this time including the BPT-AGN.
This test will be of limited use since BPT-AGN galaxies may not be similar in luminosity nor in temperature to those galaxies which host a low-luminosity AGN. 
We resort to this method since we are investigating the FIRC itself and so we cannot use the FIRC to distinguish low-luminosity AGN from star-forming galaxies.

\section{Results \& Discussion}\label{sec:results}
\subsection{Isothermal fits}\label{sec:results:fits}
\begin{figure}
  \includegraphics[width=\linewidth]{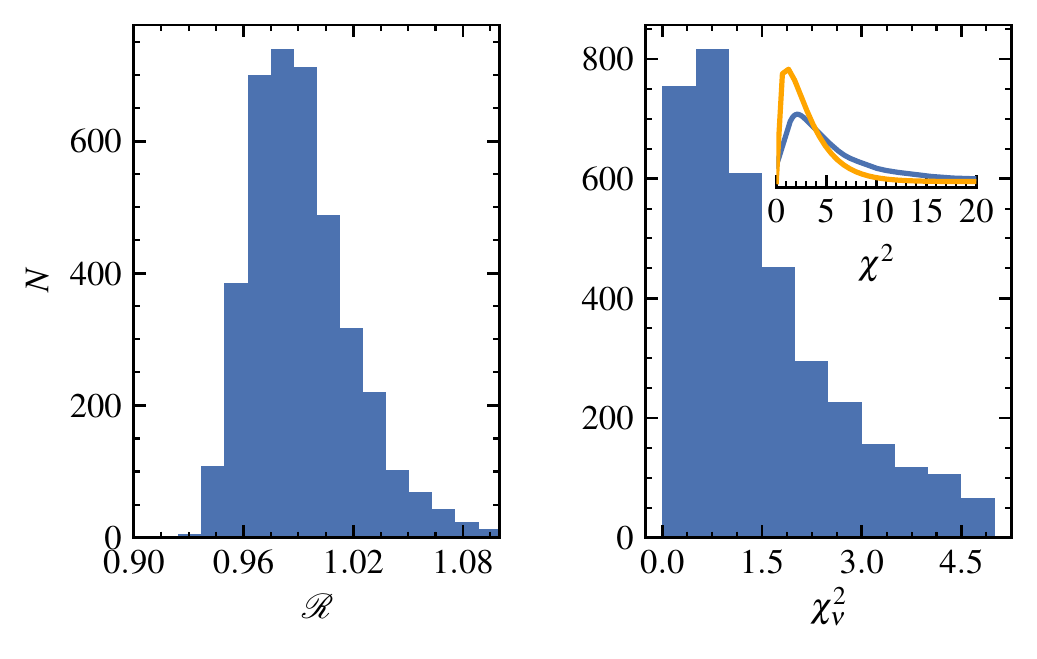}
  \caption{Fit diagnostics for our full star-forming sample. 
  The Gelman-Rubin convergence statistic histogram is shown on the left indicating that all of our fits have converged.
  The reduced $\chi^2$ distribution of the sample is shown on the right as the blue histogram. 
  The distribution of $\chi^2$ is also shown in the inset in blue, along with the $\chi^2$ distribution expected for 3 degrees of freedom for comparison in orange.
  }
  \label{fig:convergence}
\end{figure}

Before proceeding to investigate the variation of the FIRC with redshift and other parameters, we undertake several checks to ensure that our temperature estimates and \kcorrection s are reliable.
As a means of testing goodness-of-fit, we calculate the Gelman-Rubin $\mathcal{R}$ statistic for the sampled temperature and reduced $\chi^2$ for each object. 
Figure~\ref{fig:convergence} shows the distributions of $\mathcal{R}$ and reduced $\chi^2$ for our full sample of star-forming galaxies.

An $\mathcal{R} \approx 1$ signifies that all chains are sampling from the same distribution and have therefore converged \citep[see][for a full description]{Gelman1992Inference}; 
all sources in our sample have $0.9 < \mathcal{R} < 1.1$ indicating that the fits have converged.

The $\chi^2$ distribution of our sample, which we fit by least squares regression, has 3 degrees of freedom consistent with our 1-parameter model (normalisation is not fit and is instead optimised with $\chi^2$ minimisation)  when fitting with 5 bands of far-infrared observations. 
In addition, 83 per cent of our total sample have a reduced $\chi^2 < 2$. Conducting this experiment with only those sources with reduced $\chi^2 < 2$ does not affect the conclusions presented here.

\begin{figure}
  \includegraphics[width=\linewidth]{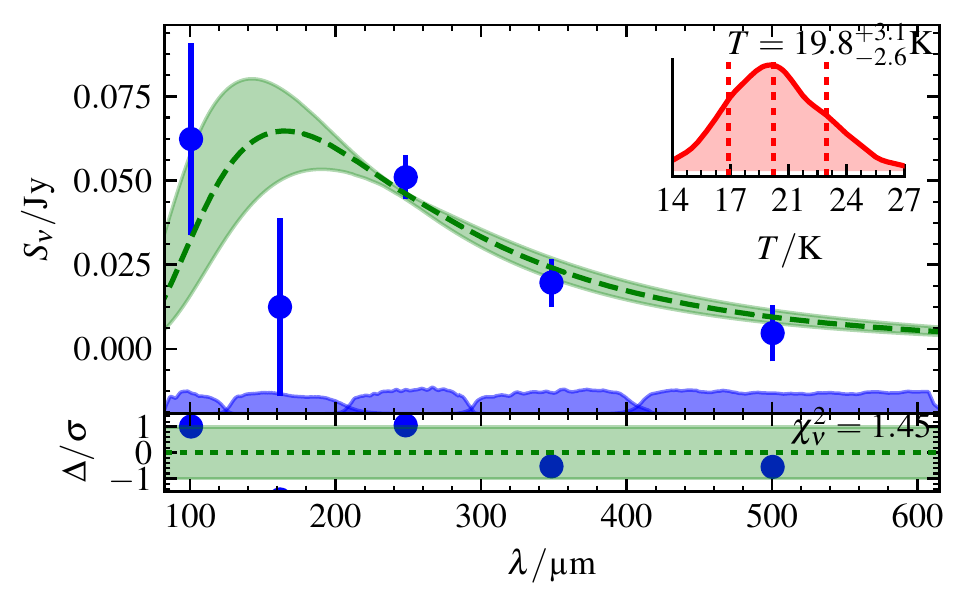}
  \caption{
  A sample isothermal fit to an SDSS star-forming galaxy at $\alpha_{J2000} = 12^{h}49^{m}46.1^{s}$, $\delta_{J2000} = 31^{\circ} 35\amin 30\asec$.
  The probability distribution for temperature is shown in the top right with the 1-sigma equal-tailed credible interval as dashed lines around the median temperature of 19.8K with a reduced $\chi^2$ of 1.45.
  The Herschel flux measurements and their uncertainties are shown as blue errorbars. 
  The fit observed-frame isothermal greybody with its own $1\sigma$ credible interval is shown as the green curve. 
  The differences between the estimated flux and the measurement are shown along the bottom axis.
  The filter transmission profiles are also shown in blue along the bottom for each wavelength.
  }\label{fig:sample_fit}
\end{figure}
\citet{Smith2013Isothermal} found that median likelihood estimators in greybody fitting are less susceptible to bias with \hatlas data than the best fit. 
Therefore, in what follows we adopt the median likelihood value from the MCMC fits as a galaxy's effective temperature for use in Equation~\ref{eq:greybody}, along with uncertainties estimated according to the 16th \& 84th percentiles of the derived distribution.
Figure~\ref{fig:sample_fit} shows an example fit. 

\begin{figure}
  \centering
  \includegraphics[width=\linewidth]{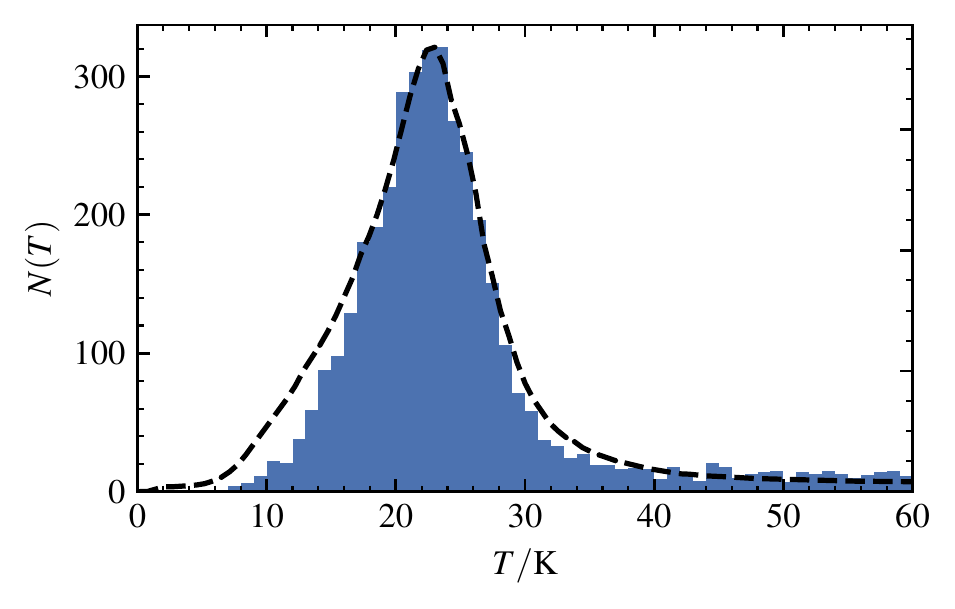}
  \caption{The distribution of median temperatures for our sample of emission-line classified star-forming galaxies (blue histogram), overlaid with the sum of temperature distributions for every galaxy obtained by MCMC (dashed line). No radio or infrared detection threshold is applied to arrive at this sample of galaxies.}
  \label{fig:temp}
\end{figure}

Figure~\ref{fig:temp} shows that our sample of emission-line classified star-forming galaxies exhibits a dust temperature distribution centred around $\sim 23$\,K with a standard deviation of $\sim 10$\,K. 
The total aggregated temperature probability distribution for all galaxies, also shown in Figure~\ref{fig:temp}, is slightly wider than the  median likelihood temperature histogram. 
This is due to the fact that the aggregated distribution includes the uncertainty from each galaxy rather than just reporting the average median likelihood temperature.

\subsection{The global FIRC at different radio frequencies}
\label{sec:results:firc}

To compare the values of \q obtained at \lofarfreq and \firstfreq, we extrapolate the \first luminosity densities to \lofarfreq assuming a power-law with a spectral index of $-0.71$.
For clarity, we label this transformed \q as $q^{\first}_{150\,\mathrm{MHz}}$ to distinguish it from the related quantity at its measured frequency, $q^{\first}_{1.4\,\mathrm{GHz}}$. 
Though it isn't especially instructive due to the large range of redshifts included in our study, we find an average value of $q^{\first}_{1.4\,\mathrm{GHz}} = 2.30 \pm 0.04$ (which is equivalent to $q^{\first}_{150\,\mathrm{MHz}} = 1.61 \pm 0.04$) which is consistent with previous studies \citep{Ivison2010FarInfraredRadio,Smith2014Temperature} to within $1 \sigma$. 
We find that the average FIRC is not consistent between low and high radio frequencies, with $q^{\lofar}_{150\,\mathrm{MHz}} = 1.42 \pm 0.03$ and $q^{\first}_{150\,\mathrm{MHz}} = 1.61 \pm 0.04$.

These values of aggregate \q are inclusive of all our star-forming-classified sources. 
A spectral index calculated from detected sources will be unreliable and a bias towards flatter spectral indices would be introduced due to the differing sensitivities and depths of \lofar and \first.
Free-free absorption is also an issue at low frequency, where it flattens the radio SED, and so may have an effect on \q, but we do not correct for its influence here.
To check whether the difference in \q between low and high frequency is due to spectral index we find the value of $\alpha$ which allows $q^{\lofar}_{150\,\mathrm{MHz}} - q^{\first}_{150\,\mathrm{MHz}} = 0$ for sources detected at $3\sigma$ in both bands.
The value for the spectral index that we find from the mean-stacked \q of these sources is $-0.58\pm0.04$ (Gaussian distributed)  which is in agreement with \citet{Gurkan2018LofarHAtlas}.
We note that we do not use this value for the spectral index in our analysis because it will be biased by only considering the brighter sources that are $3\sigma$ detected.
Instead we continue to use the value of $-0.71$ from \citet{Mauch2013325Mhz} as originally stated. 

\begin{figure}
  \centering
  \subfigure[The FIRC as measured with \lofar at \lofarfreq]{\includegraphics[width=\linewidth]{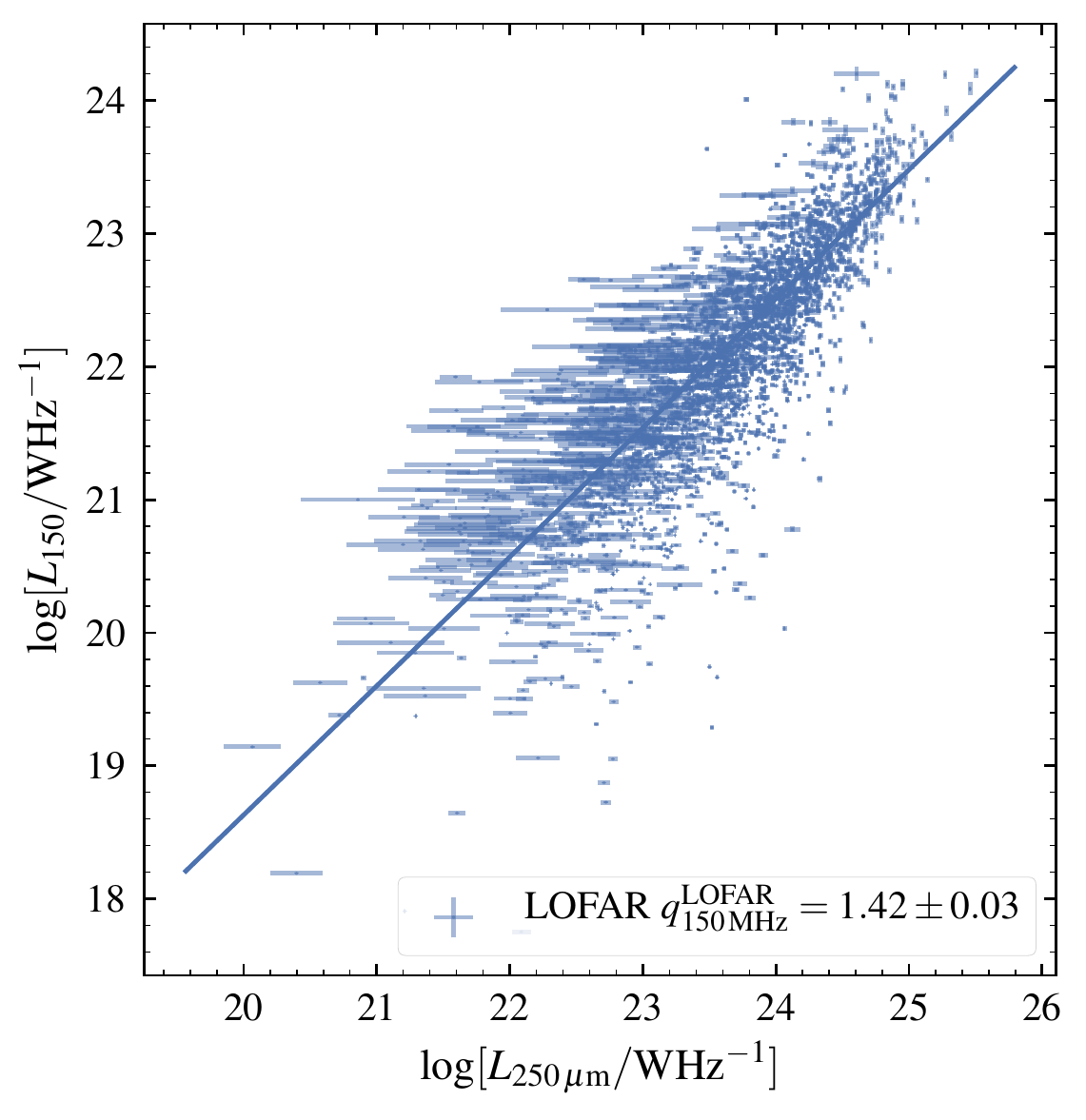}}
  \subfigure[The FIRC as measured with \first at \firstfreq]{\includegraphics[width=\linewidth]{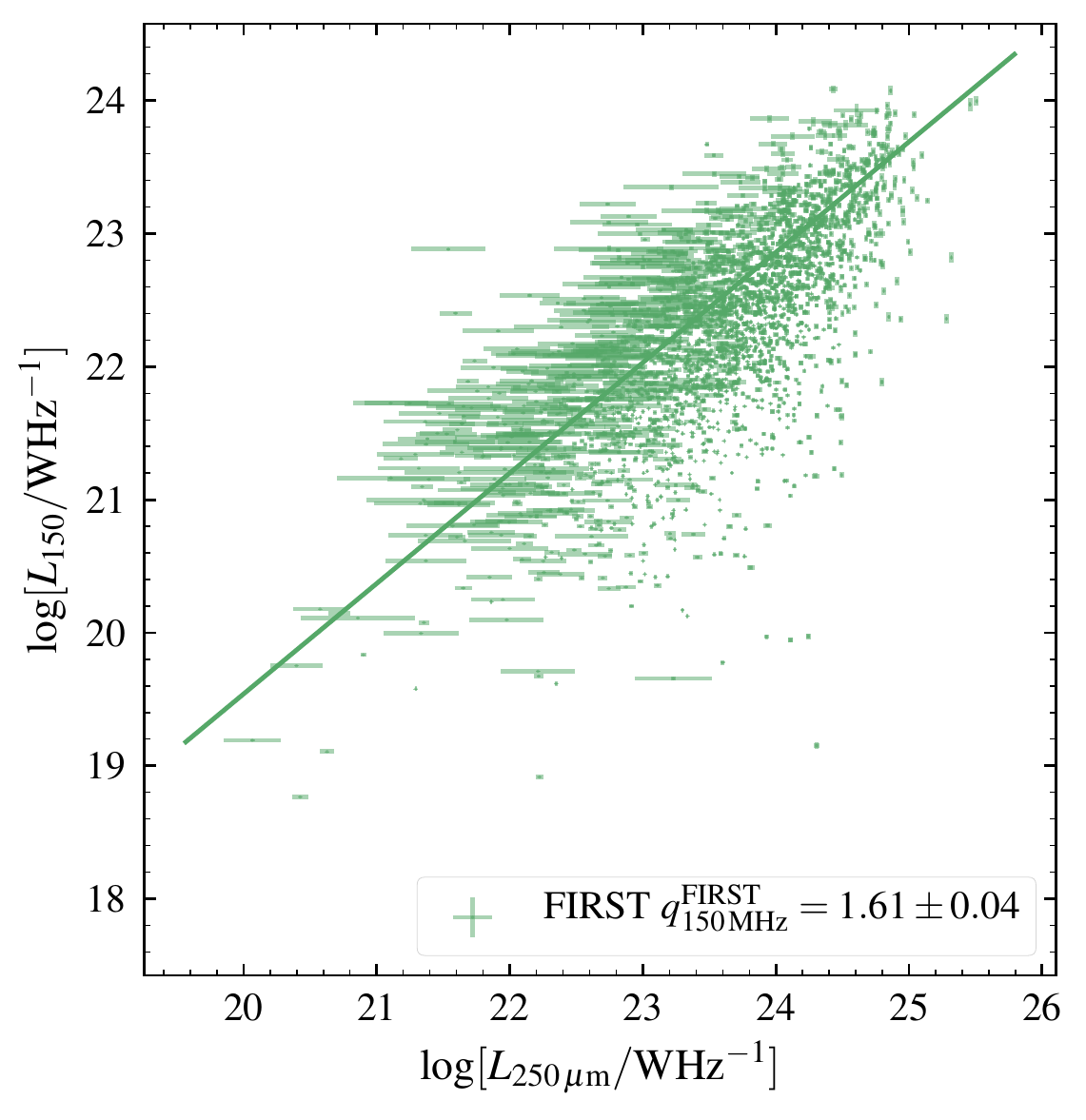}}
  \caption{The Far-Infrared Radio Correlation for \lofar(blue) and \first (green).
  The points shown are $> 3\sigma$ detected in radio and infrared fluxes, showing two clear but distinctly different correlations at \firstfreq and \lofarfreq. 
  The fit lines are power-law fits to the all sources in our star-forming sample including non-detections.
  For the purpose of comparison the \first \firstfreq luminosity densities have been transformed to 150\,MHz assuming a power law with spectfral index from \citet{Mauch2013325Mhz}.}
  \label{fig:firc}
\end{figure}
We fit the slope of the FIRC to our star-forming sample for \lofar and \first using Equation~\ref{eq:slope}.
We find that the FIRC measured with \lofar is described by $L_{150}^{\lofar} = 10^{-0.77\pm 0.19}L_{250}^{0.97\pm0.01}$ with an intrinsic width of $0.89\pm0.02$ dex. 
This is slightly below the value of unity quoted for pure calorimetry.
The FIRC measured with \first is described by $L_{150}^{\first} = 10^{2.94\pm 0.25}L_{250}^{0.83\pm0.01}$ with an intrinsic width of $1.04\pm0.03$ dex. 
We show these fits graphically in Figure~\ref{fig:firc} and include supplementary fits to the FIRC over different ranges of mid-infrared colour and specific star-forming rates in Appendix~\ref{app:supplementary}.

\subsection{The evolution of the FIRC}
\label{sec:results:evolution}
As discussed in Section~\ref{sec:intro}, there have been numerous studies of the redshift evolution of the FIRC.
\begin{figure}
  \centering
  \includegraphics[width=0.95\linewidth]{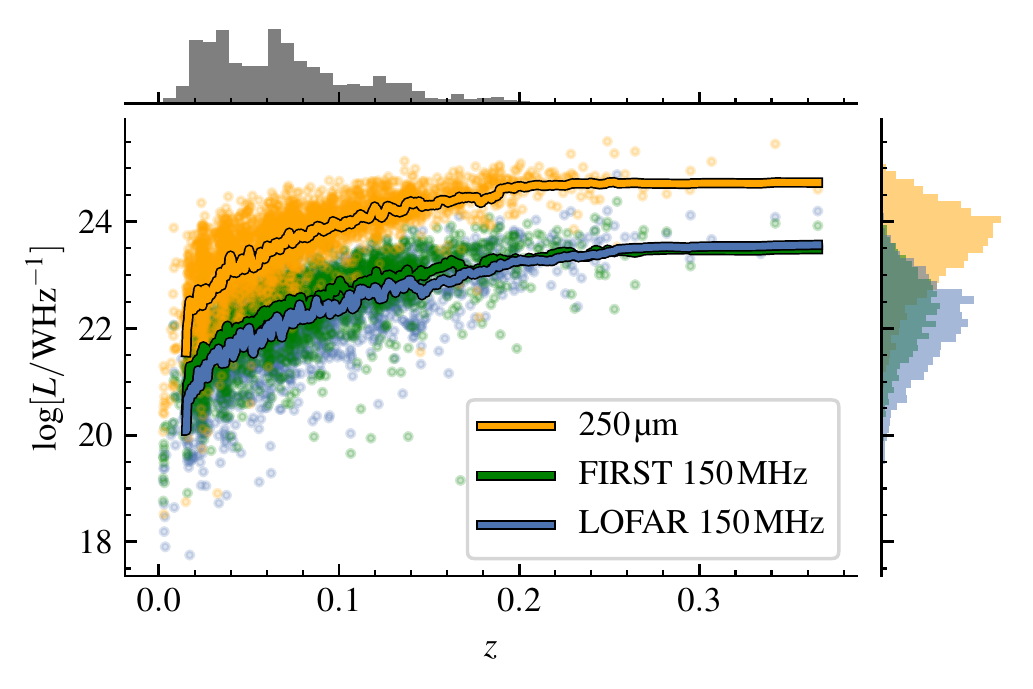}
  \caption{Distributions of Herschel SPIRE 250\,\micron\xspace (yellow), \first (green), and \lofar (blue) luminosity densities over redshift for our main star-forming sample.
  A rolling mean (inclusive of all non-detections) with a window size of 200 points is plotted to guide the eye.}\label{fig:luminosities}
\end{figure}
\begin{figure}
  \centering
  \includegraphics[width=0.95\linewidth]{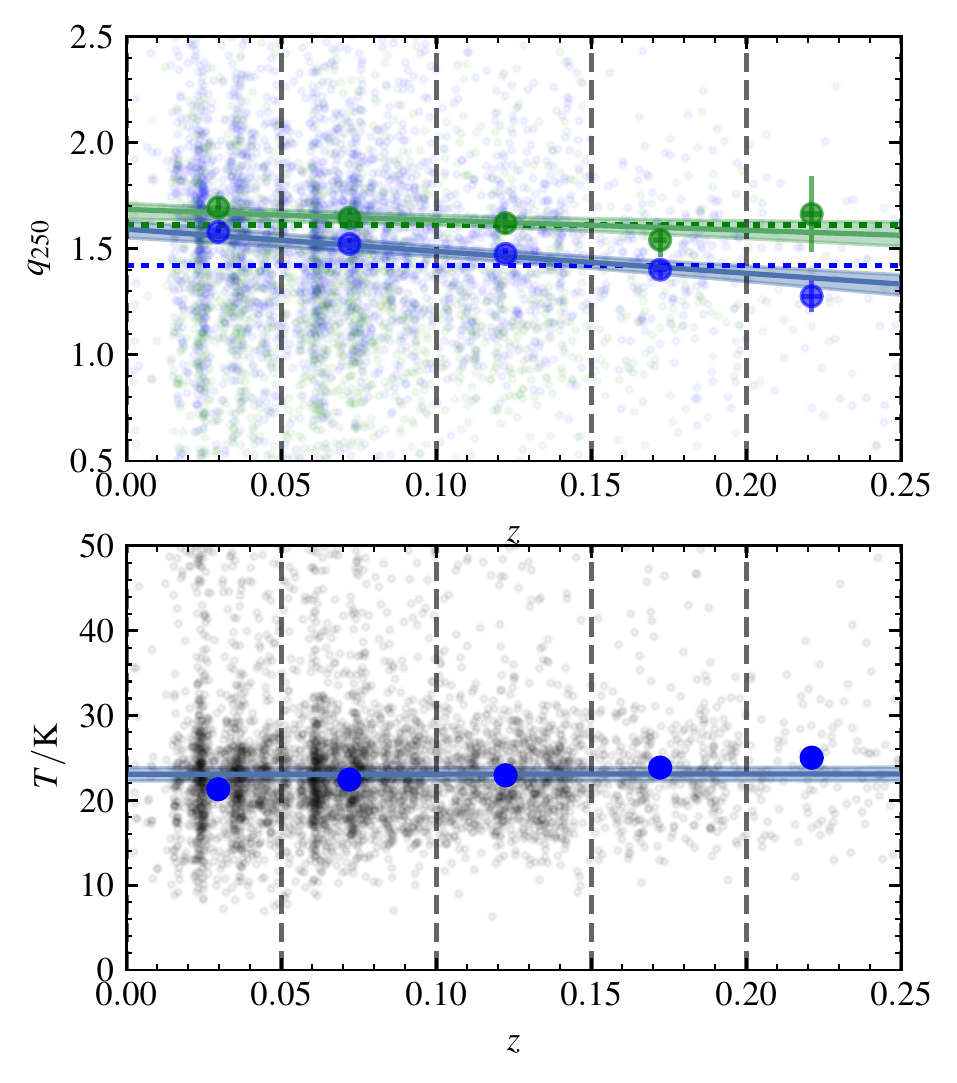}
  \caption{{\bf Top:} Evolution of $q_{250}$ over redshift measured with \lofar at 150\,MHz (blue) and \first transformed to 150\,MHz (green).
  The dashed horizontal line in the upper plot is the mean-stacked $q_{250}$ for all star-forming galaxies taken from Figure~\ref{fig:firc} for \first and \lofar at 150\,MHz.
  The coloured lines indicate the straight line fit to all galaxies in our sample binned in redshift for \lofar and \first. 
  {\bf Bottom:} The temperature in each bin, calculated by constructing an infrared SED from the average $K$-corrected flux of each source in every band and fitting Equation~\ref{eq:greybody} to the result. 
  The temperature and uncertainties are overlaid with a straight line fit to the data.
  The vertical dashed lines represent bin edges.}
  \label{fig:zevolution}
\end{figure}
Figure~\ref{fig:luminosities} shows the evolution of 250\micron~and radio luminosity densities over our redshift range for context.
To quantify the evolution of temperature and \q with redshift we fit a straight line using the Bayesian method detailed in \citet{Hogg2010Data} and implemented with \textsc{PyMC3} \citep{Salvatier2016Probabilistic}. 
We show these redshift relationships in Figure~\ref{fig:zevolution}.

To calculate the effective temperature in each bin, the \herschel fluxes are mean-stacked and their uncertainties are derived from bootstrapping.
Uncertainties on the mean redshift and mean fluxes are propagated through the MCMC fit to gain an effective temperature for each bin and its uncertainty.
The uncertainty on the mean flux is small in bins with large numbers of sources, resulting in temperature uncertainties of order 2K.
Due to significance cuts made with BPT line ratios, Figure~\ref{fig:zevolution} lacks the higher redshift galaxies present in the work of \citet{Smith2014Temperature}, hence there is a large uncertainty above $z = 0.25$ (not shown). 
However, in the bins where the uncertainty on the dust temperature is small ($< 2$K), there is no statistically significant trend with redshift, consistent with \citet{Smith2014Temperature}.
With an MCMC trace of 50,000 samples for each fit, we find strong evidence of a decrease in \q over our low redshift range  for \lofar (gradient $=-1.0^{+0.2}_{-0.3}$) but no such strong evidence of such a decline with \first (gradient $=-0.5^{+0.5}_{-0.3}$), despite being consistent with \lofar to within $1\sigma$.
It is worth noting that using the median stacking results in gradients which are consistent with the gradients calculated using the mean to within $1\sigma$. We discuss the difference between the mean and median results (and lack of impact on our results) further in Section~\ref{sec:results:agn}.
A lack of evolution seen with \first is in line with the 250\,\micron~result from \citet{Smith2014Temperature}, the 70\,\micron~result from \citet{Seymour2009Investigating}, and the 70\,\micron~and 24\,\micron~result from \citet{Sargent2010VlaCosmos}.
\citet{Rivera2017Lofar} detect an evolution at both frequencies in the Bo\"otes field and our result is consistent with theirs at redshifts below 0.25 at both frequencies.  
However, it is important to note that \citet{Rivera2017Lofar} find curved radio SEDs, suggesting that a constant slope between \lofarfreq and \firstfreq is not realistic.

At 3\,GHz, \citet{Molnar2018InfraredRadio} find no evidence for evolution in the total infrared-radio correlation in disk-dominated galaxies up until $z\sim1.5$ (though \citealt{Delhaize2017VlaCosmos} find such an evolution in $q$ using total infrared luminosity densities at redshifts $\geq 6$. 
Together with Figure~\ref{fig:zevolution}, we therefore find tentative evidence for a frequency dependence of the evolution of \q over redshift. 
However, \citet{Molnar2018InfraredRadio} also find that an evolution in \q over redshift is present in spheroids and is consistent with other studies of star-forming galaxies in general.
They suggest that AGN activity not identified with traditional diagnostics is the cause. 
Extending their conclusion to our star-forming sample may imply that the cause of the evolution found here is also low level AGN activity, with AGN prevalence increasing with redshift.

\begin{figure}
  \centering
  \includegraphics[width=\linewidth]{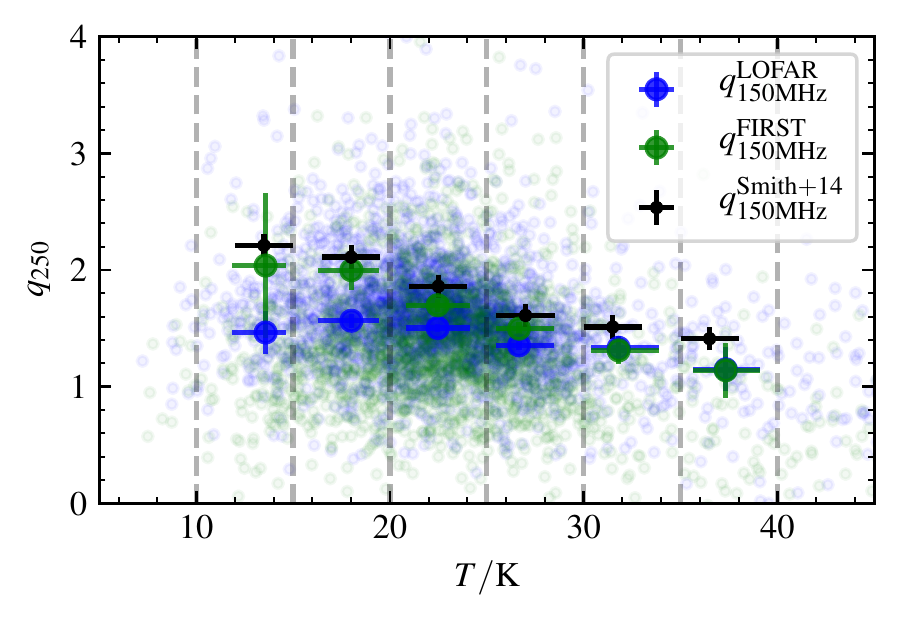}
  \caption{The temperature dependence of \q compared between high and low frequency. 
  The background dots are the individual \q calculated from the \lofar $\lofarfreq$ (blue) and \first (green) luminosity densities. 
  The \q calculated from stacked \lofar and \spire luminosity densities described earlier is plotted in bold points with errorbars derived from bootstrapping the luminosity densities within the depicted dashed bins 10,000 times. 
  The temperature uncertainties in each bin are calculated from the 16th and 84th percentiles.
  The same calculation from \citet{Smith2014Temperature} is shown as the black errorbars for comparison.}\label{fig:q_temp}
\end{figure}

Figure~\ref{fig:q_temp} shows the evolution of \q versus temperature.
For comparison, $q_{150\,\mathrm{MHz}}^{\lofar}$, $q_{150\,\mathrm{MHz}}^{\first}$ and the results of \citet{Smith2014Temperature} transformed to \lofarfreq ($q_{150\,\mathrm{MHz}}^{\mathrm{Smith+14}}$) are shown together.
Assuming a spectral index of $-0.71$, the trend of decreasing \q with increasing temperature is found with both \lofar and \first, agreeing within uncertainties when transformed to the same frequency at higher temperatures.
Cold cirrus emission is not associated with recent star-formation and so the ratio of infrared to radio luminosity (and hence $q$) will be larger for galaxies with colder integrated dust temperatures \citet{Smith2014Temperature}.
We discuss the deviation at lower temperatures in Section~\ref{sec:results:agn}.

The origin of the evolution of $q^{\lofar}_{250}$ with redshift is uncertain but we show here that the dependence of luminosity density upon redshift cannot account for all of the evolution measured in $q^{\lofar}_{250}$.
The bottom panel of Figure~\ref{fig:zevolution} shows that the average dust temperature does not depend on redshift, when averaging across the whole sample. 
Therefore, if stacked 250\,\micron\xspace luminosity density is correlated with dust temperature (and \citealt{Smith2014Temperature} show that same dependency at $250\,\micron$) in our sample, then the dependency of stacked $q^{\lofar}_{250}$ upon redshift cannot only be due to a luminosity dependence on redshift.

\subsection{Variation over the mid-infrared colour-colour diagram}\label{sec:results:mirdd}
In this section  we focus solely  on the sample of 2,901 star-forming galaxies with $5\sigma$ \wise detections in order to construct the MIRDD of \citet{Jarrett2011SpitzerWise}.
This sample covers part of the star-forming region defined by \citet{Wright2010WideField} as shown in Figure~\ref{fig:wise_sample}.
When showing \q variation of this sub-sample, we zoom in on this region.

\begin{figure}
  \centering
  \includegraphics[width=\linewidth]{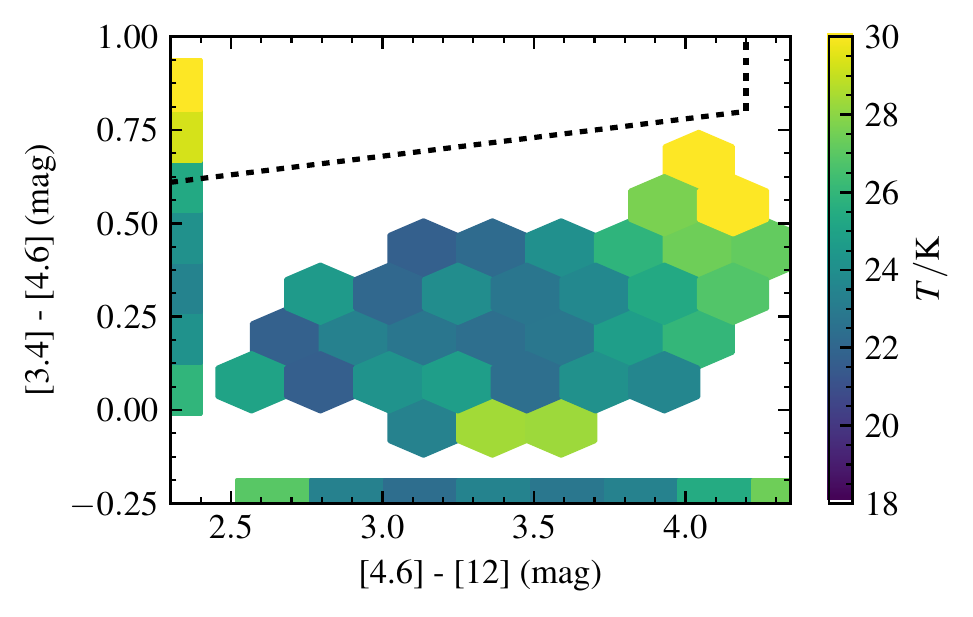}
  \caption{Mean isothermal temperature across the \citet{Jarrett2011SpitzerWise} MIRDD.
  Bins are hexagonal and are coloured linearly between 18K and 30K described by the colour bar.
  All bins have an SNR in $q^{\lofar}_{\lofarfreq} > 7$ and contain more than 50 galaxies each.
  Also plotted are the marginal bins summarising horizontal and vertical slices of the entire plane.
  These slices also obey the two conditions set on the hexagonal bins.
  For reference, the box described by \citet{Jarrett2011SpitzerWise} to contain mostly QSOs is marked by dotted lines.}
  \label{fig:temp_mirdd}
\end{figure}

\begin{figure}
  \centering
  \subfigure[\lofar at \lofarfreq]{\includegraphics[width=0.9\linewidth]{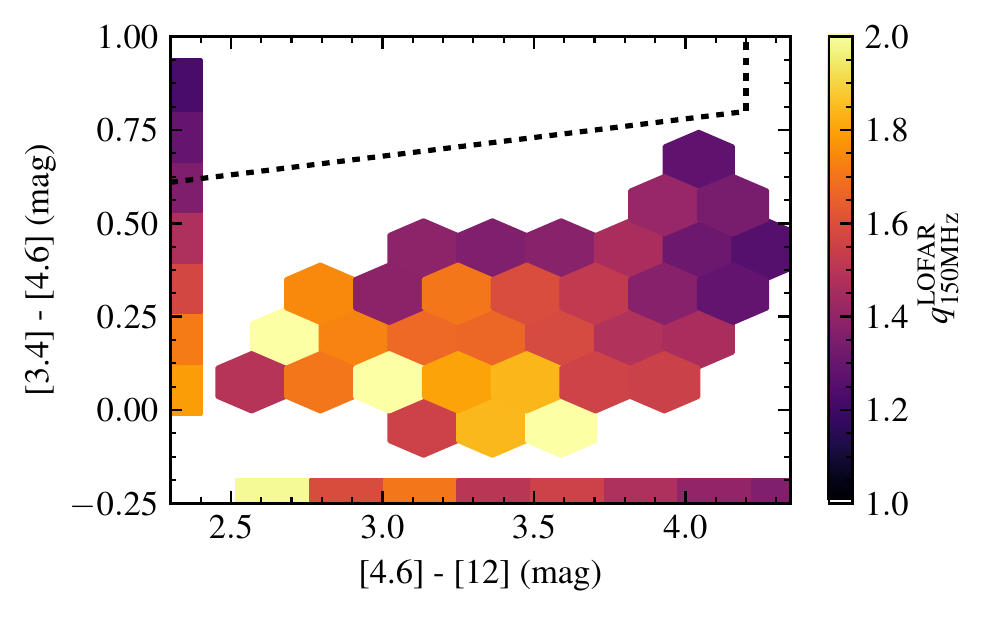}\label{fig:q_mirdd_lofar}}
  \quad
  \subfigure[\first transformed to \lofarfreq assuming $\alpha = -0.71$]{\includegraphics[width=0.9\linewidth]{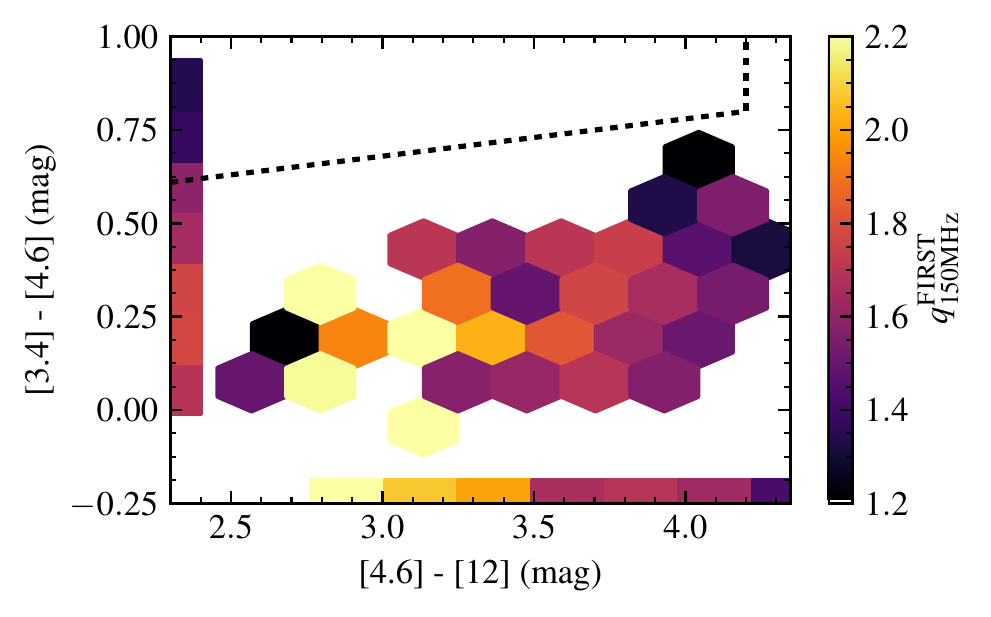}\label{fig:q_mirdd_first}}
  
  \caption{Mean-stacked \q across the \citet{Jarrett2011SpitzerWise} MIRDD.
  Bins are hexagonal and are coloured linearly according to the scale shown on the right.
  All bins have an SNR in \q$> 3$ and contain more than 50 galaxies each. 
  Also plotted are the marginal bins summarising the horizontal and vertical slices of the entire plane.
  These slices also obey the two conditions set on the hexagonal bins.
  For reference, the box described by \citet{Jarrett2011SpitzerWise} to contain mostly QSOs is marked by dotted lines.}\label{fig:q_mirdd}
\end{figure}

We calculate the mean values of temperature and \q as described in Section~\ref{sec:methods} over hexagonal bins in the \wise colour space. 
We show  only those bins which contain more than 50 galaxies and have a stacked \q with $SNR > 3$.
When these conditions are applied, 33 and 29 contiguous bins remain for \lofar and \first respectively, all with a high SNR in binned $q_{\lofarfreq}^{\lofar}$,  $q_{\lofarfreq}^{\first}$ of at least 7 and 3 respectively. 
Figure~\ref{fig:temp_mirdd} shows the mean isothermal temperature in each bin.
There is a clear and smooth increase in temperature towards redder \wisex and \wisey colours. 
The isothermal temperature of our sample increases towards the area populated mainly by starburst and Ultra-Luminous Infrared (ULIRG) galaxies. 
Our sample is positioned away from the \citet{Jarrett2011SpitzerWise} AGN area, shown as a dashed box in Figures~\ref{fig:temp_mirdd} and~\ref{fig:q_mirdd}, although we note that radiatively inefficient radio-loud AGN may populate other regions of this plot \citep{Gurkan2018LofarHAtlas}. 

The trend in temperature over mid-infrared colour is reflected in Figures~\ref{fig:temp_mirdd} and~\ref{fig:q_mirdd}, where the \q measured using both \first and \lofar decreases with  redder \wise colours in a similar fashion to temperature. 
The higher sensitivity of \lofar in comparison to \first is reflected in the much smoother relation between binned \q and mid-infrared colours.

Both the \q parameter (for both frequencies) and the temperature change smoothly across mid-infrared colour. 
We interpret this smooth variation of the temperature over \wisex colour towards more heavily star-forming galaxies as tracing the specific star formation rate of a population of normal star-forming galaxies.

To quantify the observed trend with mid-infrared colour we use a Bayesian method to find the correlation coefficients of stacked \q against both \wise colours. 
From Figure~\ref{fig:q_mirdd}, \q clearly correlates with both \wisey and \wisex.
However, since redshift is also highly correlated with \wisey and \q is independently correlated with redshift, it is necessary to control for the effects of redshift using partial correlation \citep{Baba2004Partial} in order to quantify the effect of mid-infrared colour on \q. 
We also control for isothermal temperature and stellar mass to see if all of the variation in \q over mid-infrared colour can be accounted for by covariances with those parameters. 

Our method consists of fitting a trivariate normal distribution to \wisex (x), \wisey (y), and \q to obtain correlation-coefficient estimates ($\rho_{x}$ and $\rho_{y}$). 
We estimate the correlation-coefficients for \q without controlling for any other parameters ($\rho_{x\cdot \emptyset}$ and $\rho_{y\cdot \emptyset}$) and for the residuals in \q obtained from fitting a linear relationship to \q against $z$, $T_{eff}$, and $M_{*}$.We fit the correlation coefficients with an LKJ prior  \citep{Lewandowski2009Generating} using the \textsc{PyMC3} \citep{Salvatier2016Probabilistic} model specification along with \textsc{emcee} Ensemble sampler used above. 
LKJ distributions represent uninformative priors on correlation matrices and their inclusion allows us to randomly sample correlation coefficients.

\begin{figure}
  \centering
  \subfigure[Independent partial correlation coefficient PDFs]{\includegraphics[width=\linewidth]{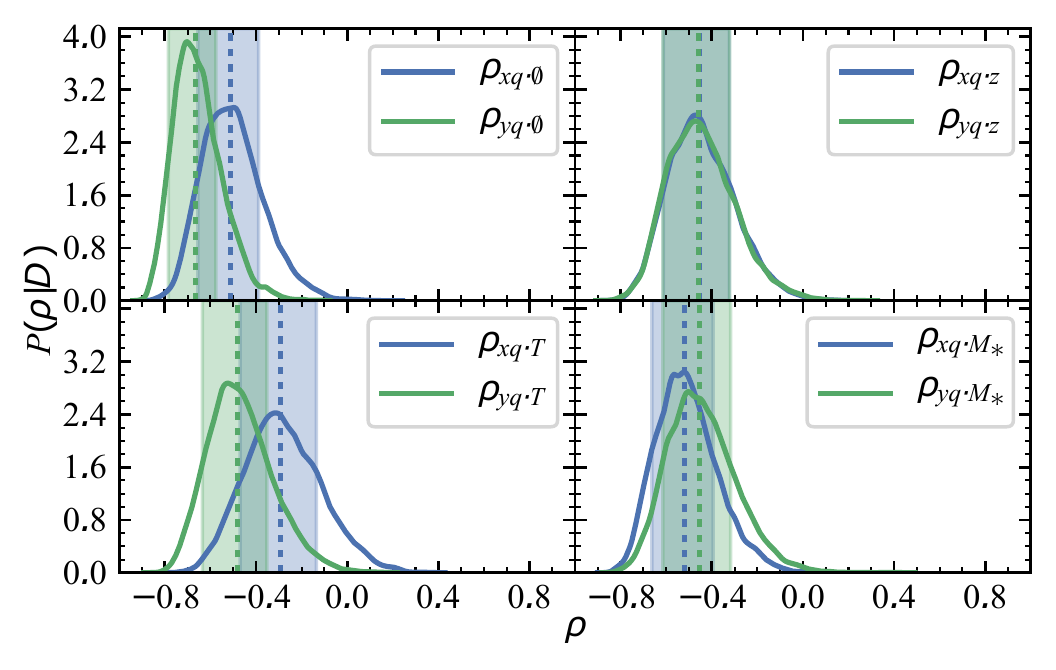}\label{fig:individual_correlation}}
  \quad
  \subfigure[Partial correlation coefficient PDF controlling for all variables at once.]{\includegraphics[width=0.8\linewidth]{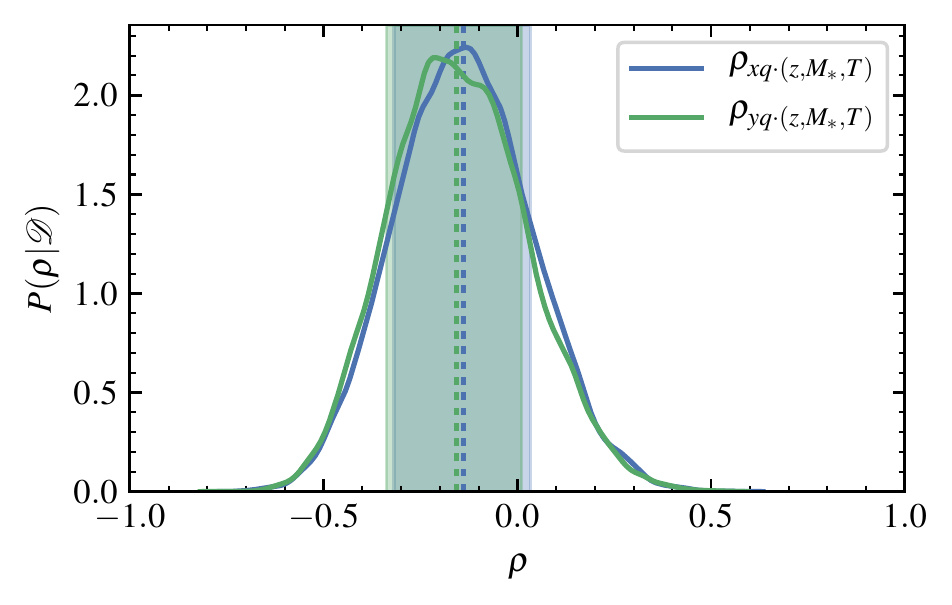}\label{fig:total_correlation}}
  \caption{
  The marginalised probability density, $P(\rho|\mathcal{D})$, distributions for the correlation coefficients ($\rho$) of \wisex (blue) and \wisey (green) against stacked $q_{250}^{\lofar}$. 
  $\rho=(-)1$ corresponds to maximal (anti-) correlation, whilst $\rho=0$ corresponds to no correlation.
  \textbf{Top left (a):} The correlation coefficient PDFs calculated assuming that $q_{250}^{\lofar}$ does not depend on other variables.
  \textbf{Top right (a):} The correlation coefficient PDFs after controlling for a linear dependence of $q_{250}^{\lofar}$ upon redshift.
  \textbf{Bottom left (a):} The correlation coefficient PDFs after controlling for a linear dependence of $q_{250}^{\lofar}$ upon effective temperature.
  \textbf{Bottom right (a):} The correlation coefficient PDFs after controlling for a linear dependence of $q_{250}^{\lofar}$ upon stellar mass.
  \textbf{Bottom panel (b):} The correlation distribution when controlling for all three parameters at once. 
  The vertical lines mark the median value for the correlation coefficient with the shaded areas marking the $16-84$th percentile range.
  A Gaussian kernel was used to smooth the probability distributions.}
\label{fig:correlation}
\end{figure}

To represent the correlation of \q over the two dimensions of \wise colour space, Figure~\ref{fig:correlation} shows the the marginalised probability distributions for each correlation coefficient.
The top panel of Figure~\ref{fig:correlation} shows the effect of controlling for redshift, temperature, and stellar mass independently as well as a naive fit which accounts for no other influential variables. 
The bottom panel of Figure~\ref{fig:correlation} shows the probability distribution of the correlation coefficients when controlling for redshift, temperature, and stellar mass at the same time. 
Initially, the distribution of \q is highly correlated with both MIR colours ($-0.5\pm 0.1$ and $-0.7\pm0.1$ for \wisex and \wisey colours respectively). 
Figure~\ref{fig:correlation} as a whole shows that the variation of \q with either \wise colour cannot be satisfactorily explained by a dependence on temperature, redshift, or stellar mass individually, but by all three at once.
This results in correlation coefficients of $0.1\pm0.2$ and $0.2\pm0.2$ for \wisex and \wisey respectively.

Using the model described above, we find that the effects of stellar mass, dust temperature, and redshift upon \q explain 16, 36, and 48 per cent of the total explainable correlation of \q over the \wisey and 8, 71, and 21 per cent over \wisex, respectively.
However, the effects of these parameters on the variation of \q are not independent of each other.
Indeed, there are non-zero covariances between these parameters, \eg, the effect of stellar mass and dust temperature upon \q at once is not equivalent to the sum of their independent effects.

Luminosity in 250\,\micron~and both radio bands increases towards redder WISE colours and hotter temperatures, consistent with evidence of a luminosity-temperature relation found by \citet{Chapman2003Median}, \citet{Hwang2010Evolution}, and in the radio by \citet{Smith2014Temperature}. Given that the temperature evolution over redshift in our sample is consistent with being flat to within the $1\sigma$, we can conclude that such a luminosity-temperature relation is not simply due to redshift effects. 
This is more evidence of the trend in \q tracing the specific star formation rate.

\begin{figure}
  \centering
  \includegraphics[width=\linewidth]{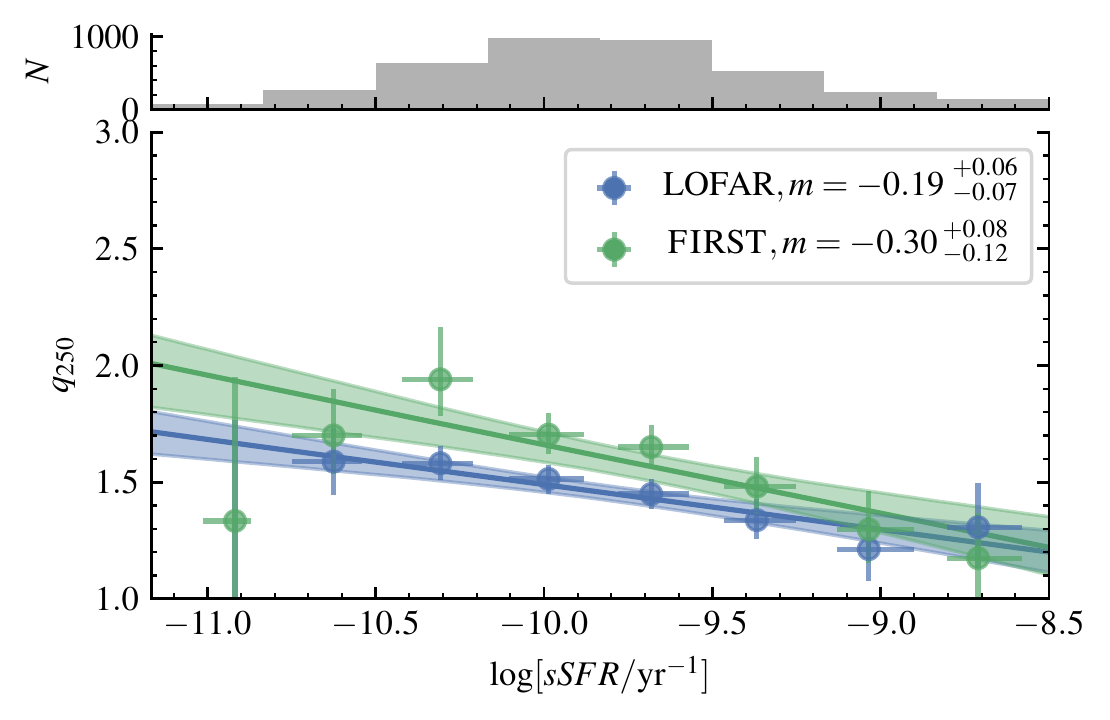}
  \caption{\q for \lofar (blue) and \first (green) at 150\,MHz against the specific star formation rate in 8 bins of width 0.3 dex.
  The uncertainties on \q are calculated via bootstrapping within each bin.
  The uncertainties on sSFR are calculated from the 16th and 84th percentiles in each bin.
  Straight line fits are shown as coloured lines with $1\sigma$ credible intervals shown as shaded regions.
  The top histogram shows the number of galaxies in each bin.
}
\label{fig:q_vs_ssfr}
\end{figure}

To test our assumption that the \wisex colour traces specific star formation rate, we use the specific star formation rates obtained from \magphys fits \citep{Smith2012HerschelAtlas}.
Figure~\ref{fig:q_vs_ssfr} shows a highly significant trend (both gradients are non-zero with a significance above $3\sigma$) between \magphys specific star formation rate and \q for both \first and \lofar (low sSFR is discussed below). 
The gradients  of the trend at high and low frequency are consistent to within $1\sigma$.

\citet{Gurkan2018LofarHAtlas} have found that above a stellar mass of $10^{10.5}M_\odot$, a strong mass dependence of radio emission, inferred to be non-AGN in origin, emerges. 
We show here that the for the variation of \q over the MIRDD to be explained, the effects of stellar mass and specific star-formation rate (for which isothermal temperature is an effective proxy) must be taken into account since they independently explain 25 and 38 per cent of the total correlation respectively.

\subsection{Potential AGN contamination}\label{sec:results:agn}
BPT classification identifies  AGN based on emission line ratios. 
However, star formation and AGN activity are not mutually exclusive \citep{Jahnke2004Uv,Trump2013Census,Rosario2013Mean} and one ionisation process can mask the other.
Indeed, the BPT diagram shows a population of Seyfert 2 objects seamlessly joined to the star-forming branch \citep{Baldwin1981Classification,Kewley2006Host}. 
Obscured AGN SEDs are bright in the mid-infrared due to the re-radiated emission from their obscuring structure \citep{Antonucci1993Unified,Stern2005MidInfrared}.
In particular, radiatively efficient  QSOs and obscured AGN are expected to be detected by WISE and to be located in the reddest space on the WISE MIRDD \citep{Jarrett2011SpitzerWise}.

\subsubsection{Searching for hidden AGN}
Whilst it may be difficult to exclude composite galaxies based purely on line ratios, spectra can be searched for AGN features and radio images inspected for signs of jets or compact cores.
The angular resolutions of \first and \lofar are too low to distinguish AGN cores from compact starbursts, but we can rule out obvious radio loud contamination.
To look for signs of physical differences between the low and high \q areas and to check for the impact of radio-mode AGN, we take two samples of galaxies.
The first sub-sample, named ``\wise-blue'', we take from the region of highest \q and bluer WISE colours. 
This region is described by the conditions $2.5 < \wisex < 3.25$ and $0.0 < \wisey < 0.4$ and so should correspond to lower-luminosity star-forming galaxies.

The second sub-sample, named ``\wise-red'', we take is described by the conditions $3.75 < \wisex < 4.5$ and $0.2 < \wisey < 0.6$, and is characterised by the lowest values of \q. 
This is the area most likely to be contaminated by AGN, given its proximity to the QSO box defined in \citet{Jarrett2011SpitzerWise} and low value of \q. 
We note that we have removed the 12 sources which lie within the QSO box before conducting the analysis here.

We visually inspected the \first and \lofar images of 100 randomly-chosen galaxies from the \wise-blue and \wise-red sub-samples for signs of cores and jets.
However, although the sub-samples are selected based on their position in the MIRDD, they also correspond to different redshift ranges.   
The higher redshift sources are selected at redder \wise colours and therefore the most luminous radio sources are selected in the \wise-red sub-sample and, conversely, the \wise-blue sub-sample consists of some of the least luminous radio sources. 
As a result, the \wise-red sub-sample tends to have extended and brighter radio emission at \lofarfreq than our \wise-blue sub-sample.
Therefore, if there is any significant AGN presence, they are more likely to populate the \wise-red sample than in the \wise-blue sample. We find little evidence of AGN activity due to compact cores or jet structures in either sample.
However, the extended emission due to the luminosity bias mentioned above makes it difficult to compare the two sub-samples.
Figure~\ref{fig:spectra} shows the rest-frame spectra, median-stacked using the method found in \citealt{Rowlands2012HAtlasGama}, for the \wise-red sub-sample in red and the \wise-blue sub-sample in blue. 
Taking the difference between the spectra of the two sub-samples indicates the potential AGN (\wise-red sub-sample) have brighter emission lines relative to their continuum and have strong $H\gamma$ and $[OIII]$ lines relative to the \wise-blue sub-sample.
The $H\gamma$ line is found to have an equivalent width of 1\,\AA\,\,which is below what would be expected for broad-line AGN \citep{Peterson1997Introduction}. Given the increased infrared luminosity, this seems to indicate that the ionisation required to excite $H\gamma$ is generated by star formation.
Moreover, if we position each median spectrum on the BPT diagram, they are both firmly within the star-forming region.

\begin{figure*}
\centering
  \subfigure[Stacked spectra for each sub-sample]{\includegraphics[width=0.4\paperwidth]{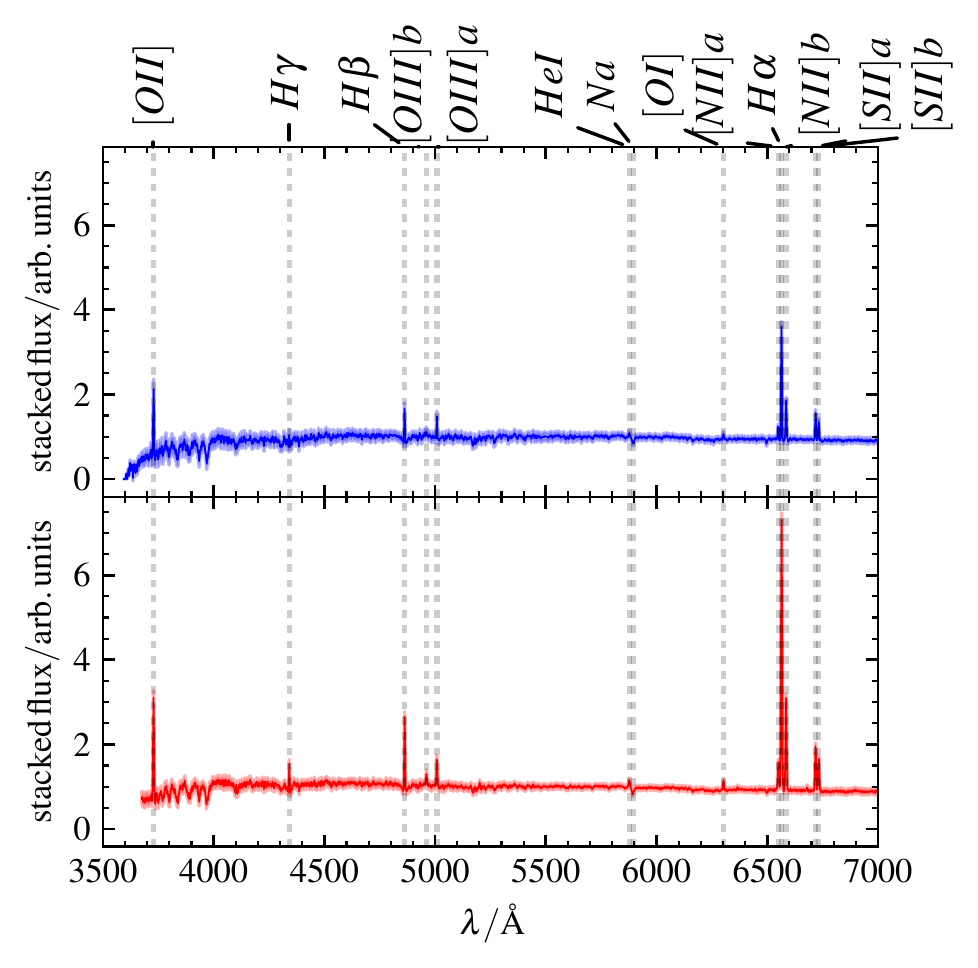}\label{fig:spectra}}
  \subfigure[Locations of the \wise sub-samples]{\includegraphics[width=0.4\paperwidth,trim={0 0 2.5cm 0},clip]{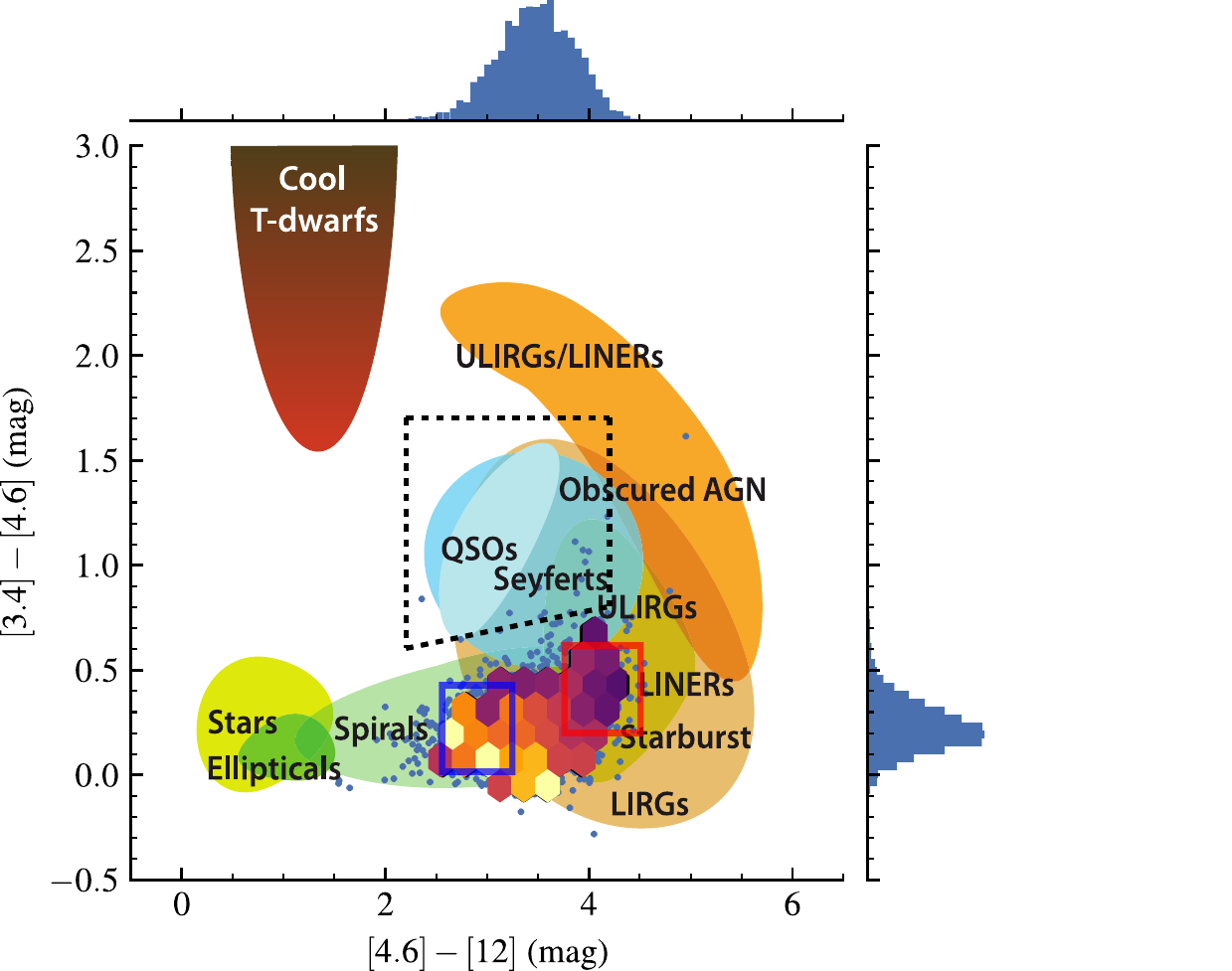}\label{fig:subsample_schema}}
  \caption{
  The median-stacked rest-frame spectra of the \wise-blue (blue) and \wise-red (red) sub-samples (left) and their locations on the MIRDD (right).
  The spectra were blueshifted to their rest frame and interpolated to a common wavelength grid between 3500\AA\,and 8000\AA.
  Each spectrum was normalised based on the median interpolated flux between 5450 and 5550\AA.
  The $1\sigma$ uncertainties in the median spectra are shown as light shaded regions around the median.
  The MIRDD is the same as in Figure~\ref{fig:wise_sample} with the binned $q_{\lofarfreq}^{\lofar}$ taken from Figure~\ref{fig:q_mirdd} overlaid.
  Blue and red boxes indicate the boundaries for the \wise-blue and \wise-red sub-samples respectively.
  }
  \label{fig:spectra}
\end{figure*}
However, a significant fraction of the radio AGN population lack characteristic emission lines \citep{Jackson1997Oa,Sadler1999Radio,Best2005Sample,Evans2006Chandra} and hence cannot be identified using a BPT classification. 
Such Low Excitation Radio AGN (LERAGN) could explain the decrease in \q with their additional contribution to radio luminosity. 
LERAGN have traditionally not been reconciled with the AGN unification model proposed by \citet{Antonucci1993Unified}. 
However, there are significant  differences between LERAGN and their standard high excitation counterparts (HERAGN) such as black hole masses \citep[\eg][]{Mclure2004Relationship,Smolcic2009Radio}, depending on sample selection \citep{Fernandes2015Black}. 
\citet{Hardcastle2006XRay} have suggested that LERAGN are the consequence of a different accretion mechanisms whereby LERAGN accrete in a radiatively inefficient mode.
LERAGN will have excess radio luminosity for their $H\alpha$ star formation, and so lie beneath the star-forming FIRC.
The radio-loud fraction of galaxies that are LERAGN or HERAGN has been found to correlate with stellar mass, colour, and star-formation rate \citep{Janssen2012Triggering}.
However, 98 per cent of our star-forming sample have stellar masses below $10^{11}\,$M$_{\sun}$, where \citet{Janssen2012Triggering} report a radio-loud fraction of LERGS below 0.001.

\begin{figure}
  \centering
  \includegraphics[width=\linewidth]{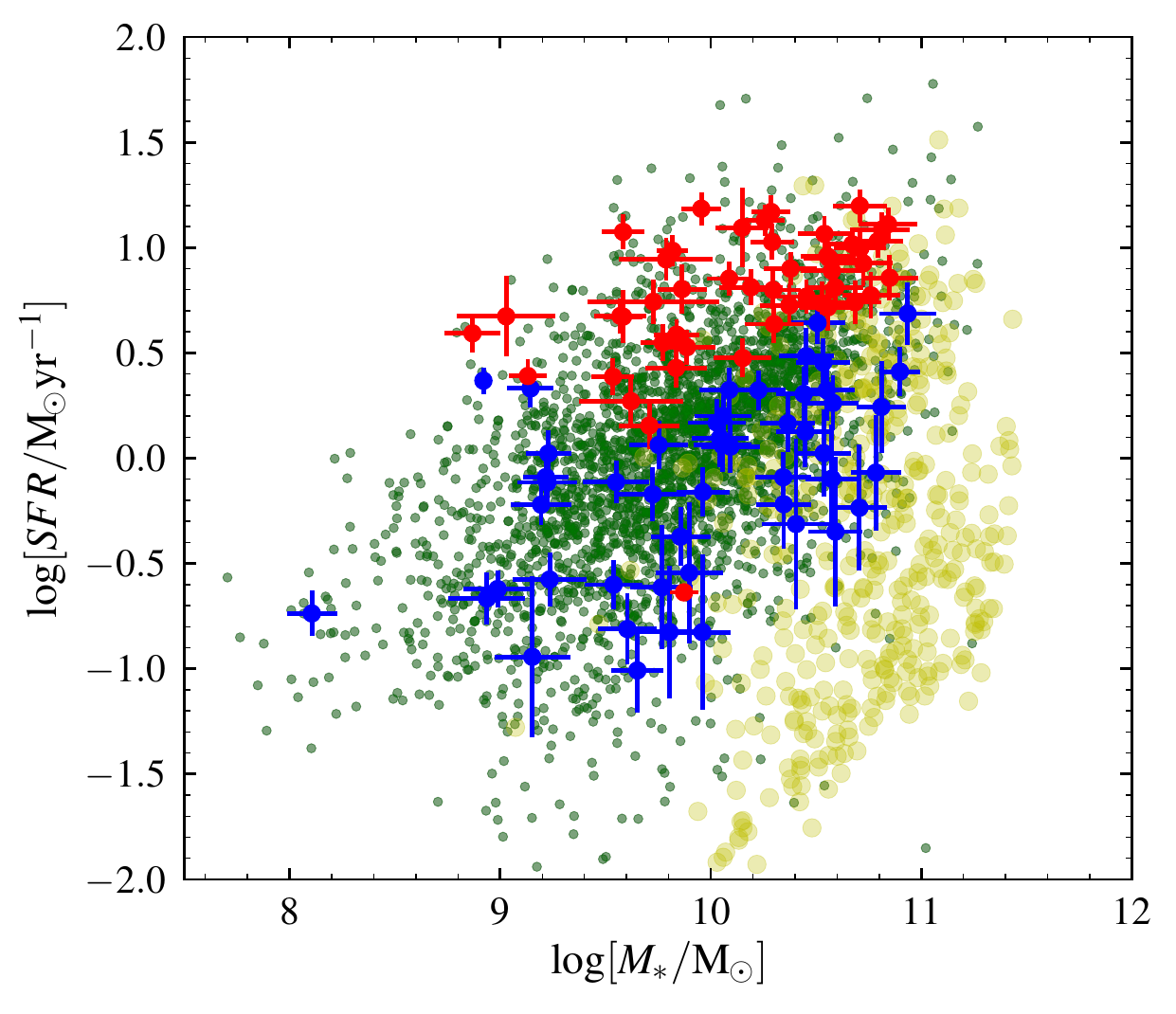}
  \caption{Stellar mass versus star formation rate for our sample.
  Our full sample is plotted in green.
  The 100 galaxies selected from both the \wise-red sub-sample (red) and the \wise-blue sub-sample (blue) are shown as the larger points with errorbars. 
  For comparison, AGN classified by BPT are plotted in yellow.
  }\label{fig:sfr_mass}
\end{figure}
In addition, \citet{Gurkan2015HerschelAtlasThe} found that radio-loud AGN tend to have lower star formation rates than star-forming galaxies, and that radio-quiet AGN also exhibit SFRs that are also offset from the star-forming galaxies.  
We use star formation rates and stellar masses fit by \magphys \citep{Cunha2008Simple,Smith2012HerschelAtlas} to test whether our sample exhibits this offset.
Figure~\ref{fig:sfr_mass} shows our entire sample including BPT-classified AGN. 
Our BPT-classified AGN, in yellow, clearly lie below the star-forming SFR-mass relation found by \citet{Gurkan2015HerschelAtlasThe} whereas our star-forming sample (in green) lie on it. 
Furthermore, our sub-sample drawn from the \wise-red region (red) lie above the rest of the galaxies suggesting again that \q decreases with increased specific star formation rate.
Figure~\ref{fig:sfr_mass} shows the galaxies within the \wise-red sub-sample (red) are representative of starbursts given their increased star formation rate. 
The fact that our BPT-classified AGN are offset in star formation and stellar mass gives some reassurance that the effect of radio-quiet AGN is minimised in our WISE sample.
Since using the mean instead of the median may well increase the effect of outliers\slash AGN, we test the effect of using the median on our results. 
We find that the discrepancy between \q measured with \lofar and \first, the evolution of \q over redshift, and the variation of \q over the MIRDD are not affected. 
In addition, we check for obvious outliers in the luminosity distributions for \lofar and \first, finding no such source.
This allows us to gain some degree of confidence that we are not plagued by outlying AGN.
However, we note that the global value of \q is increased by 0.2 dex for both \lofar and \first when using the median.

\subsubsection{Testing with the inclusion of known AGN}If we include the 447 5-sigma BPT-classified AGN and rerun our analysis, we find a large effect on both global \q values and distributions over redshift, temperature, and specific star-formation rate (which we show in Appendix~\ref{app:AGN}).
The AGN decrease the whole-sample \q to $1.23\pm 0.01$ and $1.25\pm 0.10$ for \lofar and \first respectively. 
The relative increase in radio luminosity is unsurprising and demonstrates that AGN can have a large impact on \q.

We find that the addition of BPT-AGN affect \q calculated in low temperature bins most severely  but that the difference in \q at high temperature is marginal and consistent with \q calculated with our star-forming sample.
The greatest effect is seen in the lowest temperature bin where there is a drop of 1.2 dex in \q. 
Indeed, even when we include BPT-selected AGN we find that $q_{250} - T$ trend in Figure~\ref{fig:q_temp} is still consistent with \citet{Smith2014Temperature} at high temperatures.

The largest effect that the BPT-selected AGN have on \q is found at very low specific star-formation rates.
These are likely to be LERAGN since low specific star-formation rates imply that they are quenched. 
In fact, at $\log[sSFR / \textrm{yr}^{-1}] > -10.5$ there are very few BPT-AGN; we find that a \q evaluated in this regime with AGN included is in agreement with a \q calculated with our star-forming sample. 

The evolution of \q with redshift at high and low frequency remains qualitatively the same (decreasing with redshift) with an average offset of $\approx 0.2$. 
Moreover, when the problematic low temperature and low specific star-formation rate bins are removed from the calculation, the evolution of \q over redshift at \lofarfreq and \firstfreq is unaffected.

When BPT-AGN are included in the analysis, the variation in \q over the \wise colour \wisex can be no longer be completely explained by the combined dependence on redshift, temperature, and stellar mass.
The direction of this new correlation is not towards the location of QSO box at redder \wisey colour and actually inverts the correlation direction from negative to positive (see Figure~\ref{fig:agn_correlation}, for the equivalent Figure~\ref{fig:correlation} with the addition of known AGN). 
This is likely due to the fact that we require $5\sigma$ detections in the MIRDD \wise bands and so only bright AGN are identified and positioned at bluer \wisex colours (since the 12\,\micron\, \wise band is less sensitive than 2.4\,\micron\, and 3.4\,\micron).
As a result, no BPT-AGN are found towards the reddest colours on the MIRDD (see Figure~\ref{fig:agn_mirdd}).
If such a small number of included BPT-AGN can alter the correlation in the positive direction, it is unlikely that hidden AGN are the root cause of the negative correlation found in our star-forming sample.
Since the correlation of \q with \wise colours in our star-forming sample is only just explained by these factors to $1\sigma$, this could signal either a different misclassified population of objects in our supposed star-forming sample or a feature of star-formation itself. 
However, AGN which are not detected but have a BPT classification may be positioned differently on the \wise MIRDD and have different emission properties (see discussion above). 
We therefore cannot rule out low-level AGN contamination.

Based on the above analysis, \q should only be marginally affected and therefore considered relatively trustworthy in galaxies with medium to high specific star-formation rates.

\subsection{Reconciling with star-forming models}\label{sec:results:explain}
Through the model developed by \citet{Lacki2010Physicsa}, energy loss in starbursts is mainly due to bremsstrahlung and ionisation. 
This would increase $q$ if it were not for the effect of secondary charge radiation.
Though the exact contribution of secondary charges -- resulting from cosmic ray proton collisions with ISM protons -- that is needed for a consistent FIRC is model-dependent, their addition allows the high-$\Sigma_g$ conspiracy to be maintained (\ie a linear FIRC more or less unchanging over surface density, $\Sigma_g$).
Our results show that \q decreases with specific star formation rate and hence high gas density.
Such a decrease in \q is at odds with the expected behaviour that \q should remain constant (especially at high specific star formation rates), derived from the standard model described by \citet{Lacki2010Physicsa}.
However, there are numerous reasons why the high-$\Sigma_g$ conspiracy could break down detailed by \citet{Lacki2010Physicsa}.

If the magnetic field is assumed to be dependent on volume density rather than surface density, synchrotron cooling becomes dominant and $q$ will decrease with increasing density.
The high-$\Sigma_g$ conspiracy also depends on the assumption that the escape time for cosmic rays is the same in all starbursts and normal star-forming galaxies.
If the vertical (with respect to the disk) cosmic ray diffusion scale height is constant instead, then the escape time would be two orders of magnitude smaller for starbursts than for normal star-forming galaxies.
However, we expect this effect to be much stronger than the variation in \q we see in our result and advective transport by galactic winds may dominate in spiral galaxies \citep{Heesen2018Radio}. 
Indeed, the fact that the variation is smooth and can be adequately explained by a combination of three parameters, for normal and highly star-forming galaxies, suggests that the same mass/temperature-dependent mechanism is responsible.

This work is based on the monochromatic \q and is therefore not sampling all of the reprocessed light from recent star formation. 
It is therefore possible that a FIRC based on integrated infrared luminosity does not vary as the FIRC at 250\,\micron~does. 
In a two-component model of the dust SED \citep{Charlot2000Simple}, a warm but low-mass stellar birth cloud will outshine a cold but more massive ISM at 250\,\micron. 
If the warm stellar birth clouds are more dominant at higher isothermal temperature, then the FIRC will be a less accurate calibrator of star formation rate at lower temperatures, assuming a direct relation between synchrotron and recent star formation.
Some of the variation of \q with star formation rate could then be attributed to the effect of not using integrated dust luminosities.

There are many effects to consider when modelling the FIRC and the conspiracies listed in \citet{Lacki2010Physicsa} depend on a subset of models.
We cannot say precisely which effect will reconcile their standard model with this result.

\section{Conclusions}\label{sec:conclusions}
We have used a catalogue of optically selected, BPT-classified star-forming galaxies from \citet{Gurkan2018LofarHAtlas} to study variation in the far-infrared radio correlation over redshift and other parameters.
We calculate the monochromatic far-infrared radio correlation, parametrised as \q, for \lofarfreq and compare it to that found for \firstfreq, using forced aperture photometry.
We obtained the photometry (fluxes were measured using 10 arcsec radius circular apertures centred on the optical positions) for all of these sources -- including those which are not formal detections at \lofar, \first, and \herschel wavelengths.
To avoid introducing bias to our findings, we make no significance cuts on infrared or radio fluxes.

Knowing about possible variation in the FIRC is of great importance, since a constant FIRC underpins the use of radio luminosity estimates as a star formation rate indicator. Our main results are summarised as follows:

\begin{itemize}
  \item \q at \firstfreq for our sample is found to be consistent with previous studies \citep{Jarvis2010HerschelAtlas,Ivison2010FarInfraredRadio,Smith2014Temperature}. 

  \item The FIRC for \lofarfreq is found not to be consistent with that for \firstfreq assuming a standard power law with spectral index of $-0.71$ (0.1 dex lower).  

  \item We find evidence for a decreasing \q with redshift at \lofarfreq (gradient of $-1.0^{+0.2}_{-0.3}$). 
  By comparing to the results of \citet{Molnar2018InfraredRadio}, we also find tentative evidence that the slope of this evolution becomes shallower with increasing frequency.
  An increase in radio luminosity of star-forming galaxies with redshift will be useful for high-redshift SFG detection, assuming that this evolution is maintained above $z=0.5$, as has also been suggested by FIRC studies conducted with \lofar at higher redshifts \citep[\eg][]{Rivera2017Lofar}. 

  \item We corroborate the \q-temperature variation discovered by \mbox{\citet{Smith2014Temperature}} at high frequency.
  We find that this relation also applies at low frequency to within $1\sigma$, but only at temperatures above 20K.

  \item We find that \q varies across a two-dimensional mid-infrared colour-colour space, at both radio frequencies, and within the star-forming region defined by \citet{Jarrett2011SpitzerWise}. 
  By using a hierarchical correlation model, we find that all of the correlation between \q with \wisex and \wisey colours can be attributed to the combined effects of the correlations that we measure between \q and stellar mass, redshift, and isothermal temperature, to within $1\sigma$.
  We note that the variation is not explained by redshift, temperature, or stellar mass alone but by all three in conjunction. 

  \item Using the indicative locations of different galaxy types within the WISE colour-colour plot from \citet{Jarrett2011SpitzerWise} -- \eg spirals etc -- we see that the trend to lower \q appears to reflect the transition from spirals to LIRGs to starbursts.
  \q decreases with redder \wisex colour and with increasing specific star formation rate. 
  Indeed, the lowest values of \q are seen in the region of the MIRDD occupied by the \citet{Polletta2007Spectral} starburst templates.
  Moreover, the region where LIRGs overlap with normal spirals ($3 < [4.6] - [12] < 4$) is the region where the largest gradient in \q (relative to WISE colour) is seen.

  \item To test the possible influence of AGN contamination on our results, we re-ran our analysis but this time included the BPT-classified AGN. 
  The only significant change in our results was at the lowest dust temperatures, and lowest specific star formation rates; the other regions of parameter space, and therefore our conclusions, are unchanged. 
  We can be confident, therefore, that our results are robust to the inclusion of detectable AGN, and it is tempting to attribute this variation to hitherto unknown physics of the FIRC. 
  However, we cannot totally rule out the possibility that widespread low-level AGN have some influence (though we see no evidence of high ionisation and\slash or broad emission lines indicative of their presence in stacked rest-frame optical spectroscopy for subsets of our BPT-classified SFG sample). 
  We also test for residual AGN contamination by analysing the radio images for \wise-red and \wise-blue sub-samples, finding no clear evidence for obvious AGN jet structure in either group. 
  We also test that our choice of aggregate statistic (the mean) of the parameter \q is not affected by outliers by performing the same analysis with the median.
  We find that the trends that we report of \q over redshift, sSFR, temperature, and mid-infrared colours remain unaffected by the choice of aggregate statistic with only the global value of \q changing. 
\end{itemize}

Taken together, these results indicate that the monochromatic FIRC varies strongly across the full range of BPT-classified star-forming galaxies in a manner dependent upon their mid-infrared colours (which are widely used as an empirical probe of galaxies' star formation properties), even at fixed redshift. 

We do not draw conclusions from our results alone about the efficacy of using the FIRC to calibrate radio star-formation rates, however \citet{Gurkan2018LofarHAtlas} used the same sample of galaxies, along with a full analysis of energy-balance derived stellar mass and star formation rate estimates, to investigate the low frequency radio luminosity star-formation rate relation directly.
The broken power law relation between SFR and 150\,MHz luminosity found in that work -- which they suggest may indicate the presence of an additional mechanism for the generation of radio-emitting cosmic rays -- is consistent with the possibility of residual low-level AGN contamination, and the FIRC behaviour we observe at low specific star formation rates. 
Indeed, this suggests that calibrations such as those proposed in \citet{Brown2017Calibration} may need to be more nuanced than they currently are. 

Though our results underline the exquisite combined power of \herschel  and \lofar for studying star-forming galaxies (and in particular the high quality of the maps produced by the LoTSS pipeline), it will be of great interest to investigate the star-formation and AGN content of galaxies in more detail with even more sensitive, high resolution data in the coming years, as we enter the era of the Square Kilometre Array.




\section{Acknowledgements}
We would like to thank Rainer Beck, Gianfranco De Zotti, Michal Michalowski, and Donna Rodgers-Lee for their useful comments. 
SCR acknowledges support from the UK Science and Technology Facilities Council [ST/N504105/1].
MJH and WLW acknowledge support from the UK Science and Technology Facilities Council [ST/M001008/1].
KJD acknowledges support from the ERC Advanced Investigator programme NewClusters 321271.
HJAR and GCR acknowledge the support from the European Research Council under the European Unions Seventh Framework Programme (FP/2007-2013) /ERC Advanced Grant NEWCLUSTERS-321271.
JS is grateful for support from the UK STFC via grant [ST/M001229/1].
EB acknowledges support from the UK STFC via grant [ST/M001008/1].

This work makes use of data products from \herschel-ATLAS, the Low Frequency Array, \lofar, \sdss DR7, and the Wide-field Infrared Survey Explorer, WISE.

\hatlas is a project with Herschel, which is an ESA space observatory with science instruments provided by European-led Principal Investigator consortia and with important participation from the National Aeronautics and Space Administration (NASA).
The H-ATLAS website is \url{http://www.h-atlas.org/}.

\lofar was designed and constructed by ASTRON, has facilities in several countries, that are owned by various parties (each with their own funding sources), and that are collectively operated by the International \lofar Telescope (ILT) foundation under a joint scientific policy.
The \lofar website is \url{http://www.lofar.org/}.

Funding for the SDSS and SDSS-II has been provided by the Alfred P. Sloan Foundation, the Participating Institutions, The National Science Foundation, the US Department of Energy, the NASA, the Japanese Monbukagakusho, the Max Planck Society, and the Higher Education Funding Council for England.
The SDSS website is \url{http://www.sdss.org/}.

WISE is a joint project of the University of California, Los Angeles, and the Jet Propulsion Laboratory/California Institute of Technology, funded by the National Aeronautics and Space Administration.
The WISE website is \url{http://wise.ssl.berkeley.edu/}.

This research made use of Astropy, a community-developed core Python package for Astronomy \citep{Collaboration2013Astropy}. 
The Astropy website is \url{http://www.astropy.org/}.

\bibliographystyle{mnras}\bibliography{Remote}

\appendix
\section{FIRC relations with BPT-AGN}\label{app:AGN}
Further to the discussion in Section~\ref{sec:results:agn}, we present versions of Figures~\ref{fig:zevolution},~\ref{fig:q_temp}, and~\ref{fig:correlation} which are generated from a sample of 4,541 sources formed by merging our star-forming sample with 447 BPT-classified AGN detected at $5\sigma$ in BPT emission lines. This material can be found online.

\begin{figure}
  \centering
  \includegraphics[width=\linewidth]{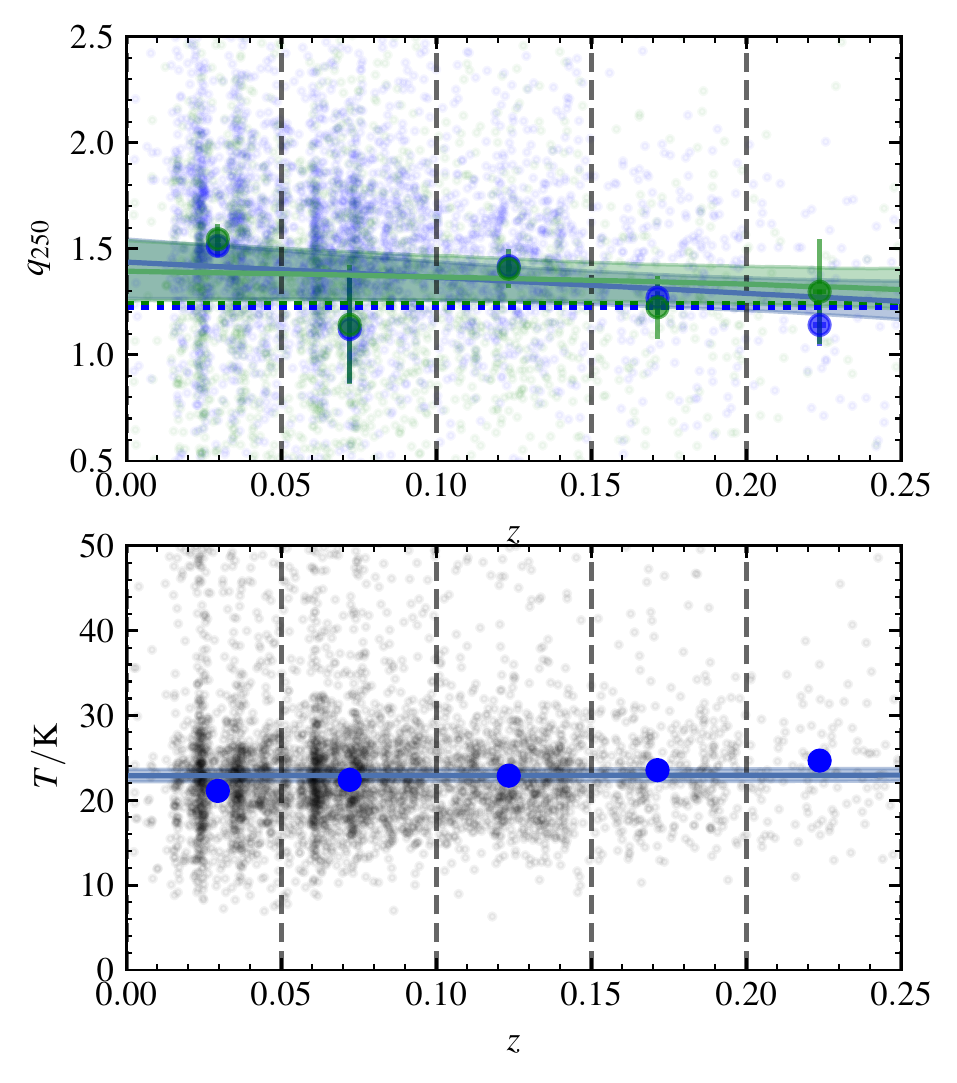}
  \caption{{\bf Top:} Evolution of $q_{250}$ over redshift measured with \lofar at 150\,MHz (blue) and \first transformed to 150\,MHz (green).
  This plot uses our star-forming sample merged with BPT-AGN whose emission lines are detected at $5\sigma$ .
  The dashed horizontal line in the upper plot is the mean-stacked $q_{250}$ taken from Figure~\ref{fig:firc} for \first and \lofar at 150\,MHz.
  The coloured lines indicate the straight line fit to all galaxies in our sample binned in redshift for \lofar and \first. 
  {\bf Bottom:} The temperature in each bin, calculated by constructing an infrared SED from the average $K$-corrected flux of each source in every band and fitting Equation~\ref{eq:greybody} to the result. 
  The temperature and uncertainties are overlaid with a straight line fit to the data.
  The vertical dashed lines represent bin edges.}
  \label{fig:zevolution_agn}
\end{figure}

\begin{figure}
  \centering
  \includegraphics[width=\linewidth]{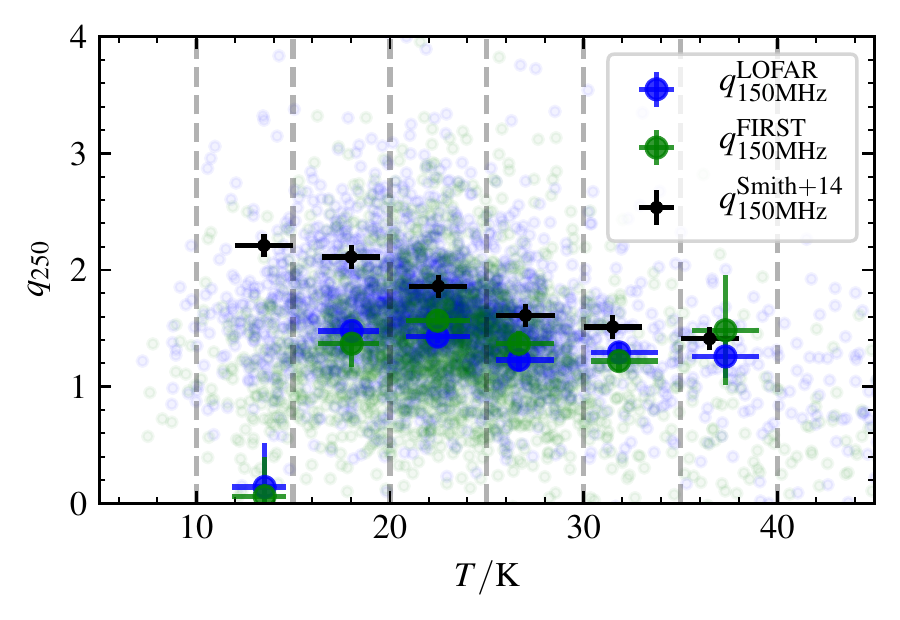}
  \caption{The temperature dependence of \q compared between high and low frequency using our star-forming sample merged with BPT-AGN whose emission lines are detected at $5\sigma$ .
  The background dots are the individual \q calculated from the \lofar $\lofarfreq$ (blue) and \first (green) luminosity densities. 
  The \q calculated from stacked \lofar and \spire luminosity densities described earlier is plotted in bold points with errorbars derived from bootstrapping the luminosity densities within the depicted dashed bins 10,000 times. 
  The temperature uncertainties in each bin are calculated from the 16th and 84th percentiles.
  The same calculation from \citet{Smith2014Temperature} is shown as the black errorbars for comparison.}\label{fig:q_temp_agn}
\end{figure}

\begin{figure}
  \centering
  \subfigure[Independent partial correlation coefficient PDFs]{\includegraphics[width=\linewidth]{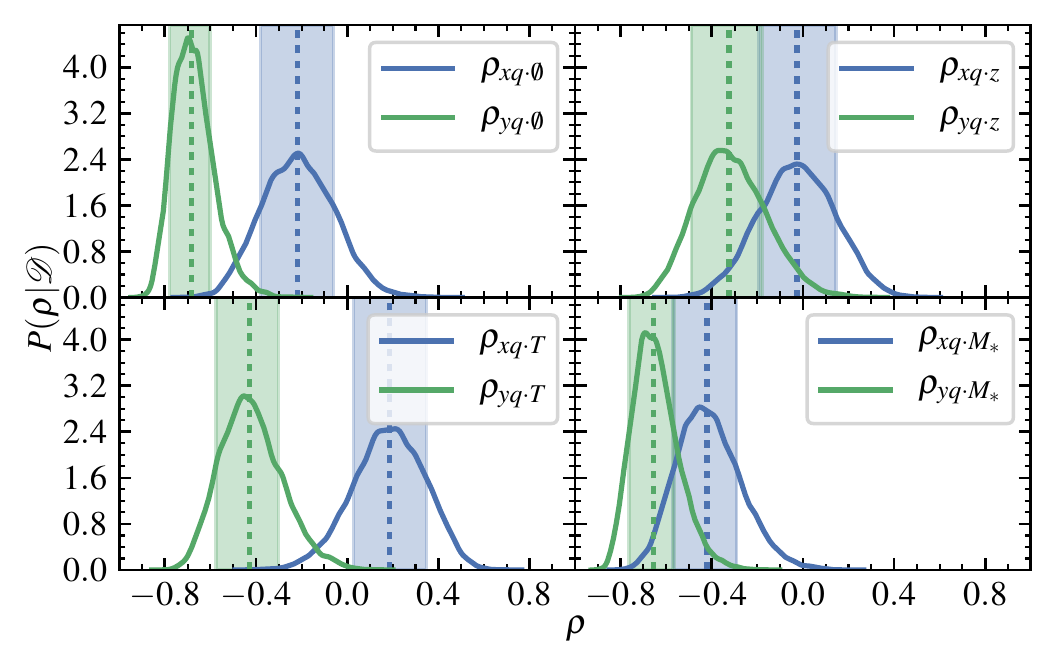}\label{fig:individual_correlation_agn}}
  \quad
  \subfigure[Partial correlation coefficient PDF controlling for all variables at once.]{\includegraphics[width=0.8\linewidth]{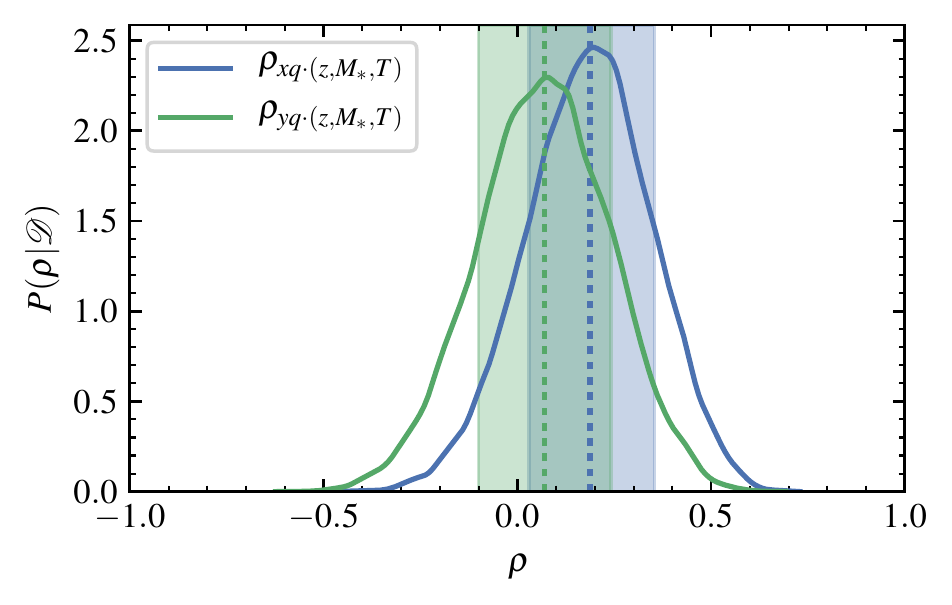}\label{fig:total_agn_correlation}}
  \caption{
  The marginalised probability density, $P(\rho|\mathcal{D})$, distributions for the correlation coefficients of \wisex (blue) and \wisey (green) against stacked $q_{250}^{\lofar}$ (using our star-forming sample merged with BPT-AGN whose emission lines are detected at $5\sigma$).
  \textbf{Top left (a):} The correlation coefficient PDFs calculated assuming that $q_{250}^{\lofar}$ does not depend on other variables.
  \textbf{Top right (a):} The correlation coefficient PDFs after controlling for a linear dependence of $q_{250}^{\lofar}$ upon redshift.
  \textbf{Bottom left (a):} The correlation coefficient PDFs after controlling for a linear dependence of $q_{250}^{\lofar}$ upon effective temperature.
  \textbf{Bottom right (a):} The correlation coefficient PDFs after controlling for a linear dependence of $q_{250}^{\lofar}$ upon stellar mass.
  \textbf{Bottom panel (b):} The correlation distribution when controlling for all three parameters at once. 
  The vertical lines mark the median value for the correlation coefficient with the shaded areas marking the $16-84$th percentile range.
  A Gaussian kernel was used to smooth the probability distributions.}
\label{fig:agn_correlation}
\end{figure}

\begin{figure}
  \centering
  \includegraphics[width=\linewidth]{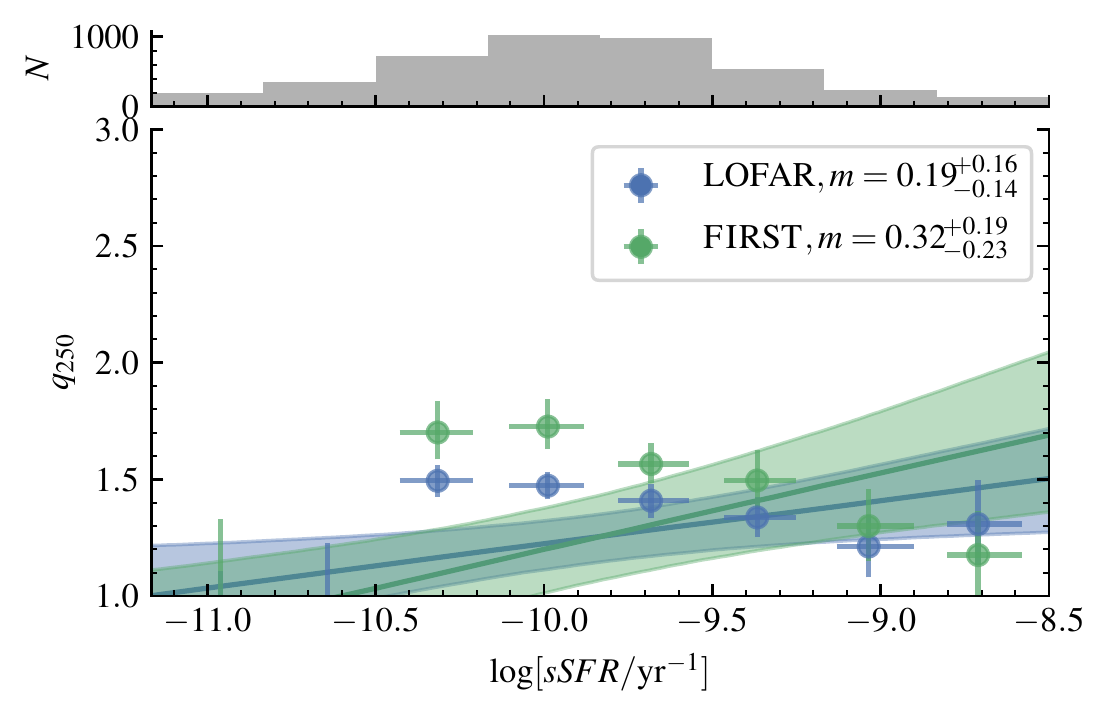}
  \caption{\q for \lofar (blue) and \first (green) at 150\,MHz against the specific star formation rate in 8 bins of width 0.3 dex.
  This plot uses our star-forming sample merged with BPT-AGN whose emission lines are detected at $5\sigma$ .
  The uncertainties on \q are calculated via bootstrapping within each bin.
  The uncertainties on sSFR are calculated from the 16th and 84th percentiles in each bin.
  Straight line fits are shown as coloured lines with $1\sigma$ credible intervals shown as shaded regions.
  The top histogram shows the number of galaxies in each bin.
}
\label{fig:q_vs_ssfr_agn}
\end{figure}

\begin{figure}
  \centering
  \includegraphics[width=\linewidth]{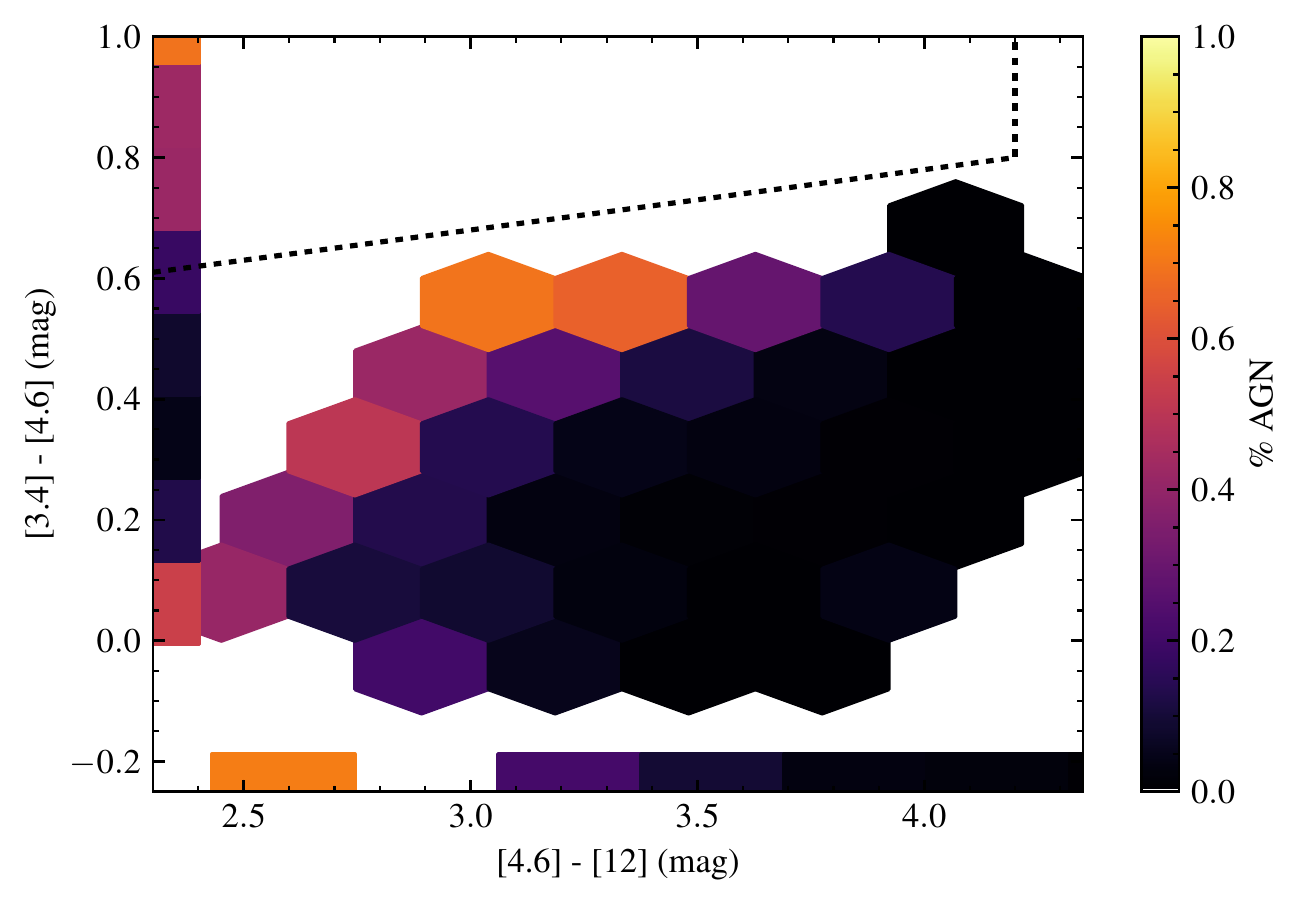}
  \caption{Mean fraction of BPT-AGN across the \citet{Jarrett2011SpitzerWise} MIRDD using our star-forming sample merged with BPT-AGN whose emission lines are detected at $5\sigma$ .
  Bins are hexagonal and are coloured linearly according to the scale shown on the right.
  All bins have an SNR in $q^{\lofar}_{250\,\micron}> 3$ and contain more than 50 galaxies each. 
  Also plotted are the marginal bins summarising the horizontal and vertical slices of the entire plane.
  These slices also obey the two conditions set on the hexagonal bins.
  For reference, the box described by \citet{Jarrett2011SpitzerWise} to contain mostly QSOs is marked by dotted lines.}\label{fig:agn_mirdd}
\end{figure}

\section{Supplementary figures}\label{app:supplementary}
This material can be found online.
\begin{figure}
  \centering
  \subfigure[\lofar at \lofarfreq]{\includegraphics[width=\linewidth]{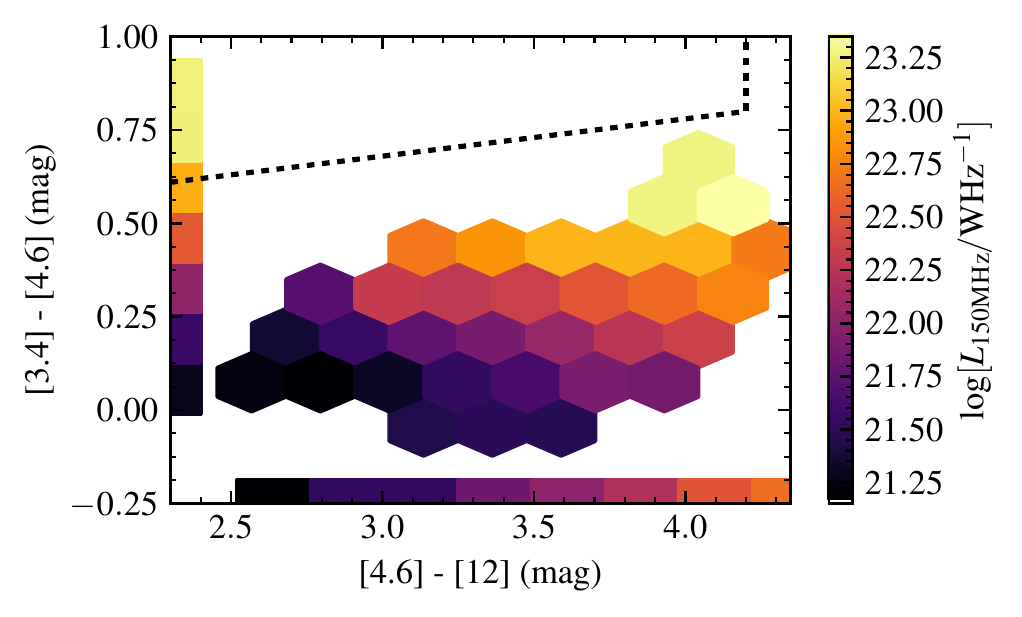}\label{fig:luminosity_mirdd_lofar}}
  \quad
  \subfigure[\first at \firstfreq]{\includegraphics[width=\linewidth]{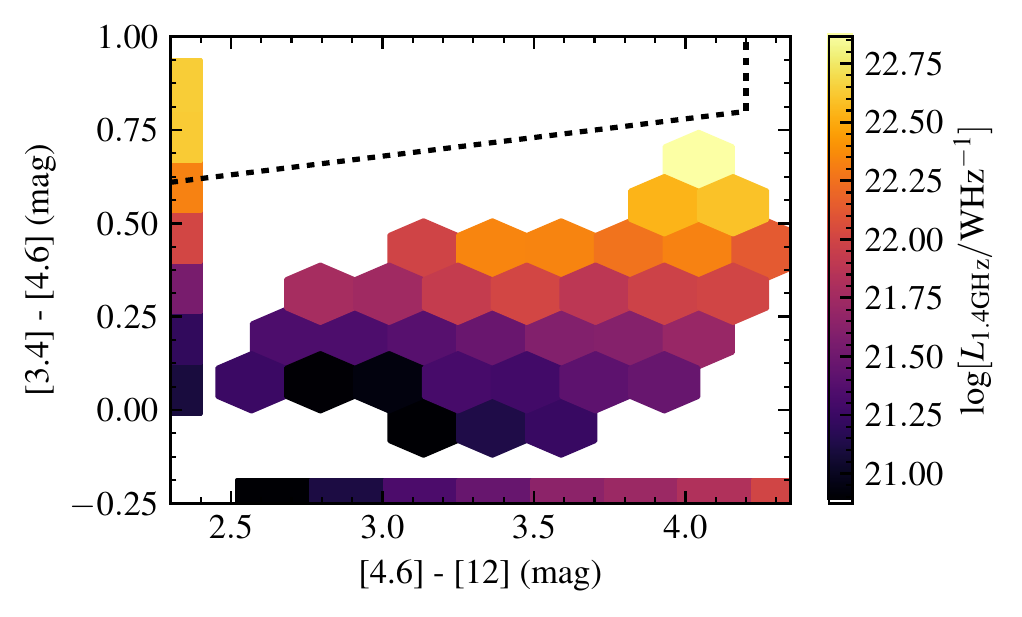}\label{fig:luminosity_mirdd_first}}
  \quad
  \subfigure[250\,\micron]{\includegraphics[width=\linewidth]{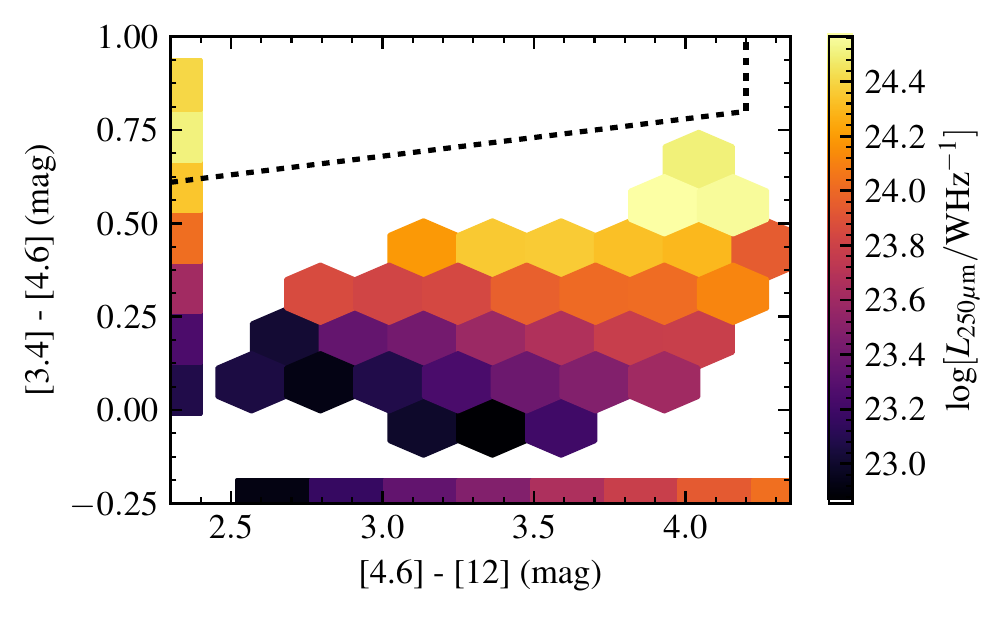}\label{fig:luminosity_mirdd_250}}
  
  \caption{Mean luminosity density across the \citet{Jarrett2011SpitzerWise} MIRDD.
  Bins are hexagonal and are coloured linearly according to the scale shown on the right.
  All bins have an SNR in $q^{\lofar}_{250\,\micron}> 3$ and contain more than 50 galaxies each. 
  Also plotted are the marginal bins summarising the horizontal and vertical slices of the entire plane.
  These slices also obey the two conditions set on the hexagonal bins.
  For reference, the box described by \citet{Jarrett2011SpitzerWise} to contain mostly QSOs is marked by dotted lines.}\label{fig:L_mirdd}
\end{figure}

\begin{figure}
  \centering
  \includegraphics[width=\linewidth]{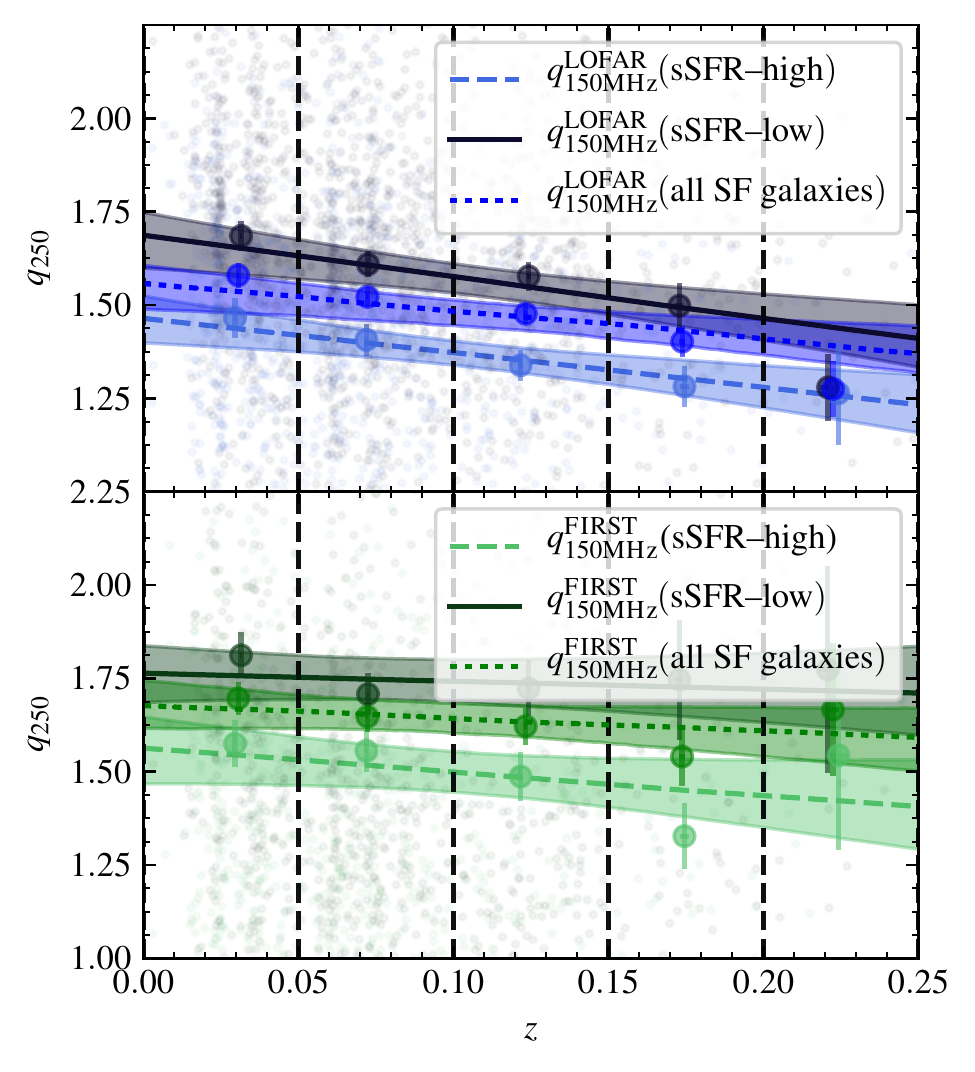}
  \caption{Evolution of $q_{250}$ over redshift for the high-sSFR (light points, dashed lines), low-sSFR (darker points, solid lines), and all star-forming galaxies (dotted line) measured with \lofar at 150\,MHz (blue) and \first transformed to 150\,MHz (green).
  The coloured lines indicate the straight line fit to all galaxies in our sample binned in redshift for \lofar and \first.}\label{fig:ssfr_high_low_qz}
\end{figure}

\begin{figure}
  \centering
  \includegraphics[width=\linewidth]{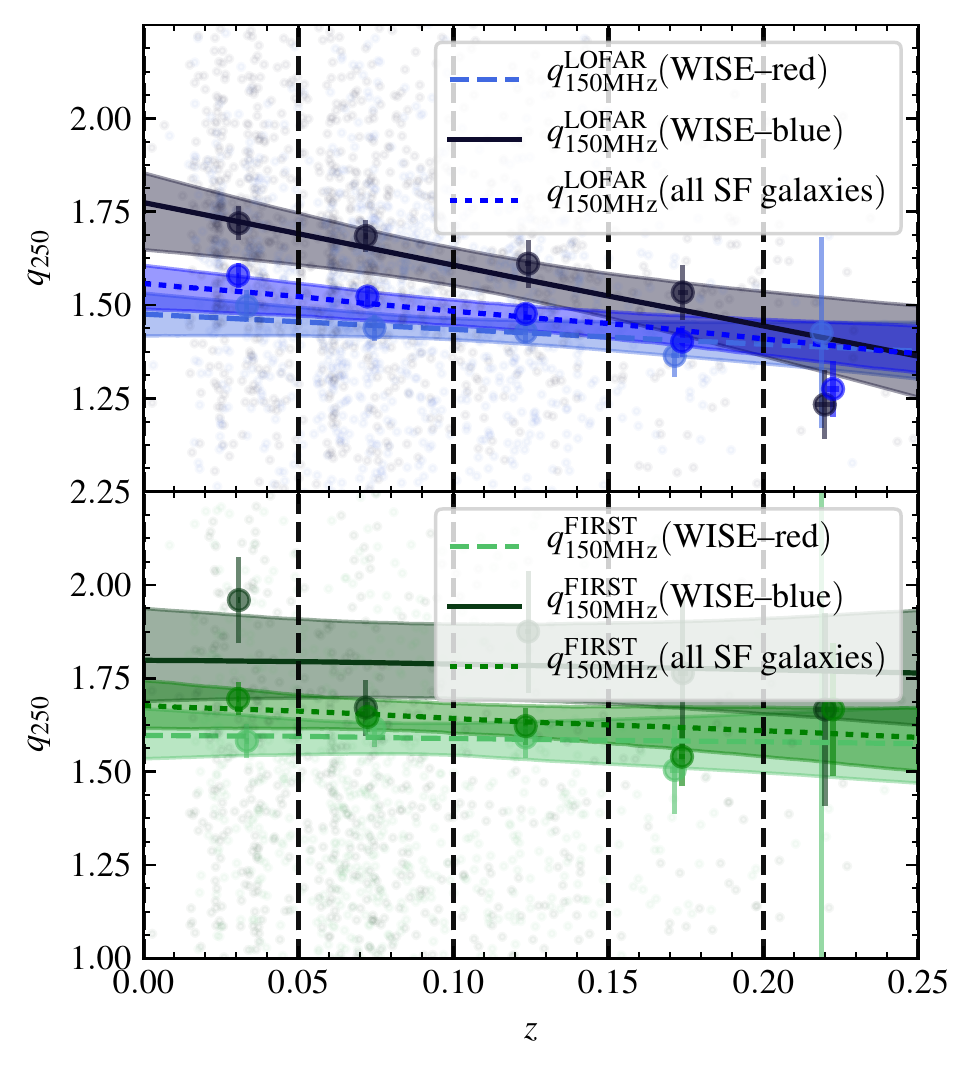}
  \caption{Evolution of $q_{250}$ over redshift for the \wise-red (light points, dashed lines), \wise-blue (darker points, solid lines), and all star-forming galaxies (dotted line) measured with \lofar at 150\,MHz (blue) and \first transformed to 150\,MHz (green).
  The coloured lines indicate the straight line fit to all galaxies in our sample binned in redshift for \lofar and \first.}\label{fig:wise_red_blue_qz}
\end{figure}

\onecolumn

\begin{figure}
  \centering
  \includegraphics[width=\linewidth]{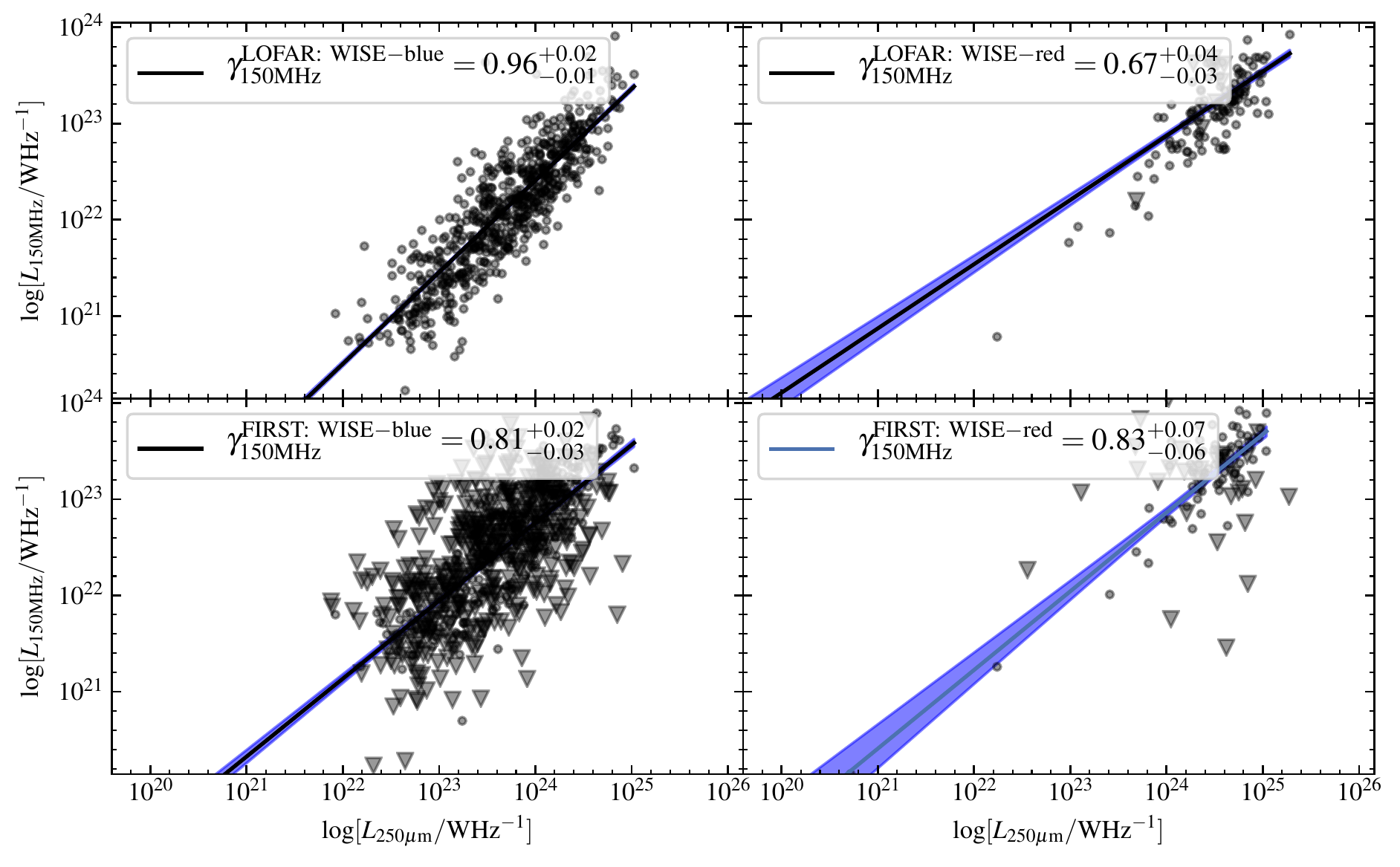}
  \caption{Fits to the FIRC measured by \lofar and \first for the \wise-red and \wise-blue sub-samples.
  $3\sigma$ detections are shown as black points. 
  $3\sigma$ upper limits for sources for which there is not formal $3\sigma$ detection are shown as black triangles
  The fit lines are power-law fits to the all sources in our star-forming sample including non-detections. 
  For the purpose of comparison the \first \firstfreq luminosity densities have been transformed to \lofarfreq assuming  a power law with spectral index from \citet{Mauch2013325Mhz}.
  }\label{fig:wise_slopes}
\end{figure}

\begin{figure}
  \centering
  \includegraphics[width=\linewidth]{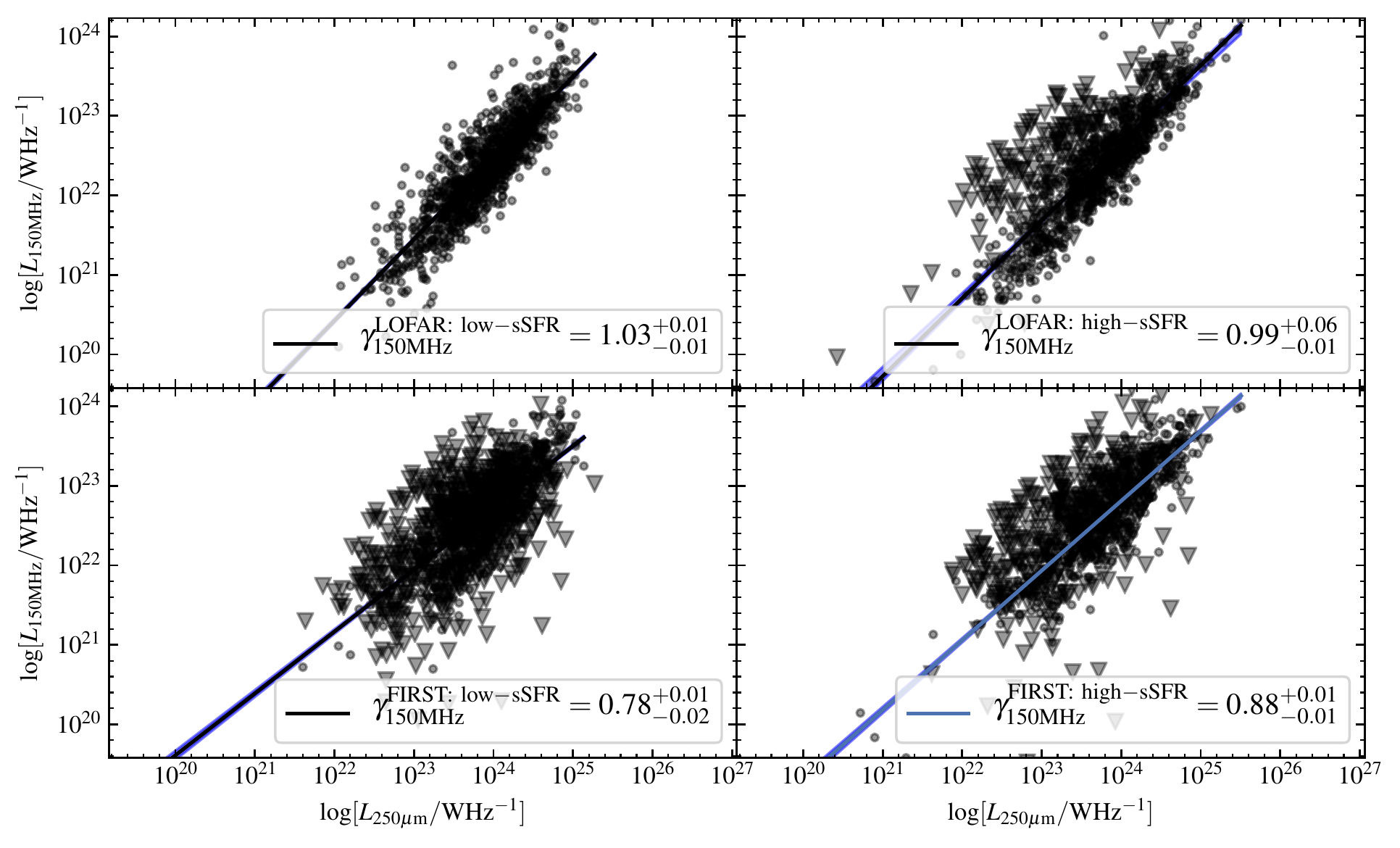}
  \caption{Fits to the FIRC measured by \lofar and \first for galaxies in our star-forming sample with \magphys fit specific star-formation rates above $-9.75$yr$^{-1}$ (high-sSFR) and below $-9.75$yr$^{-1}$ (low-sSFR).
  $3\sigma$ detections are shown as black points. 
  $3\sigma$ upper limits for sources for which there is not formal $3\sigma$ detection are shown as black triangles
  The fit lines are power-law fits to the all sources in each sample including non-detections. 
  For the purpose of comparison the \first \firstfreq luminosity densities have been transformed to \lofarfreq assuming  a power law with spectral index from \citet{Mauch2013325Mhz}.
  }\label{fig:wise_slopes}
\end{figure}

\label{lastpage}

\end{document}